\documentclass[a4paper,11pt]{article}
\usepackage{jheppub}
\usepackage[T1]{fontenc}
\usepackage{graphicx}
\usepackage{amsmath}
\usepackage{amsmath}
\usepackage{xspace}
\usepackage{color}
\usepackage[utf8]{inputenc}
\usepackage{commath}
\usepackage{slashed}
\usepackage{fancyvrb}
\usepackage{braket}
\usepackage{multirow}
\usepackage{color,soul}

\title{Sensitivity of Future Tritium\\ Decay Experiments to New Physics}

\author[a]{James A. L. Canning,}
\author[a]{Frank F. Deppisch,}
\author[a,b]{Wenna Pei}
\affiliation[a]{University College London, Gower Street, London WC1E 6BT, UK}
\affiliation[b]{Institute of Physics, ELTE Eötvös Loránd University, Pázmány Péter sétány 1/A, H-1117\newline Budapest, Hungary
}
\emailAdd{james.canning.20@ucl.ac.uk}
\emailAdd{f.deppisch@ucl.ac.uk}
\emailAdd{wenna.pei.20@ucl.ac.uk}

\abstract{Tritium beta-decay is the most promising approach to measure the absolute masses of active light neutrinos in the laboratory and in a model-independent fashion. The development of Cyclotron Radiation Emission Spectroscopy techniques and the use of atomic tritium has the potential to improve the current limits by an order of magnitude in future experiments. In this paper, we analyse the potential sensitivity of such future searches to keV-mass sterile neutrinos and exotic interactions of either the active or sterile neutrinos. We calculate the relevant decay distributions in both energy and angle of the emitted electron with respect to a potential polarisation of the tritium, including the interference with the Standard Model case as well as incorporating relevant final state corrections for atomic tritium. We present projected sensitivities on the active-sterile neutrino mixing and effective coupling constants of exotic currents, demonstrating the potential to probe New Physics in tritium experiments.
}

\begin{document}
\maketitle
\flushbottom
\section{Introduction} 
\label{sec:introduction}

The detection of neutrino oscillations implies that active neutrinos in the Standard Model (SM) have a small but non-zero mass \cite{PhysRevD.98.030001}. Oscillations are explained through a non-diagonal mixing between flavour and mass eigenstates and they crucially require that the three active neutrinos do not have identical (degenerate) masses. Instead, the observed oscillations are compatible with two different mass splitting scenarios. Oscillation experiments are insensitive to the absolute values of active neutrino masses which remain undetermined. The ordering of the neutrino mass eigenstates with respect to their flavour content is also still unknown. Matter effects in oscillations of solar neutrinos led to the determination that the so-called solar mass splitting $\Delta m_\text{sol}^2 > 0$, but the order in the atmospheric splitting $\Delta m_\text{atm}^2$ is unknown.

The absolute values of eV-scale neutrino masses are being constrained in single $\beta$-decay\footnote{Alternative direct probes are other decays with neutrinos in the final state such as that of the muon and tau. The resulting constraints are much weaker and not relevant for our discussion.} as well as in indirect measurements using neutrinoless double $\beta$-decay \cite{Agostini:2022zub, Deppisch:2016rox} and cosmology \cite{Ade:2015xua}. The process of single $\beta$-decay, in which $d\to u+e+\bar\nu$ at the quark level, will be the primary focus of this paper. In particular, we have chosen to focus on tritium $\beta$-decay as future leading experiments are planning to use tritium due to its short half-life and low $Q$-value. This decay produces helium with the emission of an electron anti-neutrino and an electron,
\begin{align}
	{}^3\text{H}\to {}^3\text{He} + e^- +\bar\nu_e.
\end{align}
In order to derive the Lagrangian we start at the quark level which (in terms of the up and down quarks, $u$ and $d$, respectively) at low energies takes the form in the SM,
\begin{align}
    \label{eq:left}
    \mathcal{L} = 
	-\frac{G_F}{\sqrt{2}}V_{ud}
	\left[\bar e\gamma^\mu(1-\gamma^5)\nu_e\right]
        \left[\bar u\gamma_\mu (1-\gamma^5)d\right] 
	+ \text{h.c.},
\end{align}
with $V_{ud}$ the CKM mixing matrix element between the up and down quarks and $G_F$ the Fermi constant. This is an effective dimension-6 operator that serves well as an approximation to the SM because we are looking at interactions with energies much below the mass of the $W$ boson.  

The advantage of single $\beta$-decay as a neutrino mass measurement technique is that it is model-independent and relies purely upon the kinematics of the decay. Specifically, it is independent of the Majorana or Dirac nature of neutrinos and as a laboratory measurement it does not rely on assumptions beyond SM particle physics, e.g., the exact form of the cosmological model in astrophysical observations. While the emitted neutrino interacts too weakly to be consistently detected, the electron is easily observed as a charged particle. The presence of the neutrino's mass will affect the emission spectrum of the electron, most clearly at the endpoint where the electron takes near the maximum energy that it can kinematically receive, leaving the neutrino with negligible kinetic energy. The challenge with this method is that only a very small fraction of the emitted electrons will have an energy near the endpoint requiring a high number of decay events to reach the required statistics. Additionally, the electrons' energies have to be measured with a very high precision. This is necessary to be sensitive to the spectral shape near the endpoint. There are many effects such as the energy uncertainties introduced through excited final states, especially when using molecular tritium, and the experimental challenges which limit the precision of such a measurement~\cite{Otten:2008zz}.

Single $\beta$-decay produces an electron anti-neutrino, which, in turn is described in terms of a weighted sum over its constituent mass eigenstates. With experiments not able to resolve the individual three thresholds of the mass eigenstates in the foreseeable future, experimental results are thus often presented in terms of an effective $\beta$-decay mass~\cite{Kleesiek2019},
\begin{align} \label{eq:mbeta}
    m_\beta^2 = \sum_{i=1}^3 |U_{ei}|^2 m_i^2,
\end{align}
see the discussion of Eq.~\eqref{eq:dGammadE-mbeta} and Fig.~\ref{fig:Kurie} below. Here, $U_{ei}$ are the elements of the Pontecorvo–Maki–Nakagawa–Sakata (PMNS) mixing matrix and $m_i$ are the neutrino masses in the respective ordering. The current strictest bound on $m_\beta$ comes from the KATRIN experiment with $m_\beta < 800$~meV at 90\% confidence level (CL) \cite{Aker2022}. KATRIN aims to reach a sensitivity of $m_\beta \sim 200$~meV (90\% CL) by the conclusion of its operation~\cite{Aker2022}. Using the experimental techniques employed in KATRIN, it will be very challenging to further improve on this sensitivity. The minimum value of $m_\beta \sim 9$~meV is the `worst'-case of normally ordered neutrino states with a massless lightest neutrino, see Fig.~\ref{fig:Kurie}.\footnote{The lower limit on $m_\beta$ is derived in the SM with three active neutrinos and a unitary PMNS matrix. It may be further modified under the presence of sterile neutrinos or other non-unitary effects.} To achieve this ultimate goal, new techniques are required and the most promising approach is based on measuring the electron energy through the emitted cyclotron radiation in a magnetic field as championed by the Project~8 collaboration~\cite{Pettus:2017sxd}. Together with utilising atomic instead of molecular tritium to minimise final state corrections, the projected sensitivity is $m_\beta \sim 40$~meV~\cite{Pettus:2017sxd}. This will guarantee the observation of $m_\beta$ for inversely ordered neutrinos with a minimum value of $m_\beta \sim 50$~meV. Exploring the potential to achieve even better sensitivities using atomic tritium and quantum technologies, the Cyclotron Radiation Emission Spectroscopy Demonstrator Apparatus is being built to develop the necessary techniques~\cite{QTNM:2020}.

Probing $m_\beta$ at the level of $\mathcal{O}(10)$~meV will require observing a large number, $N_{1\text{eV}}\approx 10^5 - 10^6$, of tritium decay electrons with energies in the last $\approx$~eV of the endpoint. Such an experiment, owing to the high count rate, may also be designed to measure the full tritium decay spectrum. Techniques required to produce atomic tritium may also allow the partial polarisation of the tritium atom, giving rise to a potentially non-isotropic angular distribution of the emitted electrons. In this work, we explore the sensitivity of such experiments to New Physics specifically, exotic charged currents beyond the SM $V-A$ term in Eq.~\eqref{eq:left}, as well as the presence of a keV-scale sterile neutrino that is either mixing with the active neutrinos, is participating in exotic currents beyond $V-A$ or both.

Such extensions are motivated by the presence of neutrino masses that necessarily require the existence of a corresponding mechanism of mass generation. Within the SM, all the fermions exist either within an $SU(2)$ left-handed doublet or a right-handed singlet, with a Yukawa Higgs coupling the two, generating a Dirac mass. A right-handed neutrino singlet would be uncharged under all SM gauge interactions and is thus called a sterile neutrino. The simplest extension to the SM is to analogously generate a Dirac mass for the neutrino. However, as this state is uncharged, there is nothing forbidding it to have a Majorana mass as well, thus making it a Majorana fermion, leading to the seesaw mechanisms for mass generation \cite{AKHMEDOV2000215}. This will induce a mixing between the SM-like active and the sterile states. An extension of the PMNS matrix of dimension $3+n_s$, where $n_s$ is the number of sterile states, will describe this mixing. In the context of tritium decay, this mixing will allow the production of sterile states as long as they are lighter than the tritium $Q$-value of $18.59$~keV. 

The inclusion of right-handed neutrinos also evokes the potential presence of a fundamental left-right symmetry within nature. We can consider the possibility that at high energies there exists an equivalent right-handed weak force which, due to some unknown mechanism, is broken at a much higher scale than the SM weak force \cite{Sirunyan2018}. This would lead to right-handed vector currents within $\beta$-decay, potentially within either or both of the hadronic and leptonic currents. Extending this further, we consider the full basis of exotic operators within the hadronic sector which could couple to either left or right-handed leptonic currents \cite{Ludl:2016ane}. 

This manuscript is organized as follows. In Sec.~\ref{Model} we describe the effective operators and associated model parameters potentially responsible for $\beta$-decay. In Sec.~\ref{Calculation} we then outline the calculation of the tritium decay rate as well as energy and angular decay distributions. In Sec.~\ref{sec:experimental} we describe the current and future experiments. In Sec.~\ref{sec:Sensitivity} we demonstrate the limits on the model parameters that could be reached by an idealized future experiment with enough statistics to probe $m_\beta \approx \mathcal{O}(10)$~meV. Sec.~\ref{sec:conclusions} concludes our analysis and the Appendix contains details of final state corrections to the decay spectrum (Appendix~\ref{App:Cor}) and full expressions of the exotic contributions considered here (Appendix~\ref{App:full-expressions}).

\section{Effective Operators for Tritium Decay} 
\label{Model}

For the decay of tritium, composed of two neutrons and one proton, a hadronic Lagrangian must be considered to take account of the internal strong dynamics~\cite{Ludl:2016ane},
\begin{align}
	\label{eq:Lag}
	\mathcal{L}_\text{SM} = 
	- \frac{G_F}{\sqrt{2}} V_{ud}(1+\delta_\beta)
	\left[\overline{e}\gamma^\mu(1-\gamma^5)\nu_e\right]
	\left[\overline{{}^3\text{He}} \gamma_\mu\left(g_V-g_A\gamma^5\right){}^3\text{H}\right],
\end{align}
with $\delta_\beta = 0$ within the SM. The coefficients $g_V$ and $g_A$ parametrize the vector and axial couplings and take the values $g_V = 1$ and $g_A\approx 1.247$, respectively \cite{Simkovic:2007yi}. It is from this Lagrangian that all of the later results denoted `SM' are derived. This form is analogous to the quark level Lagrangian. This is because the tritium and helium form an approximate $SU(2)$ isospin doublet equivalent to the quarks \cite{Simkovic:2007yi}. Further terms $\mathcal{O}(Q/m_\text{H})$ have been neglected as $Q\sim 20$~keV $\ll m_\text{H}\sim 3$~GeV, the tritium mass \cite{Ludl:2016ane}. The fermionic fields in the Lagrangian implicitly depend on their individual spin states. Rather than averaging over all of these spins, we will consider what happens when the tritium is prepared in a particular spin state. This will lead to correlations between the electron's momentum direction and the tritium's spin which will break the isotropy of emission.

\subsection{Exotic Currents}
\renewcommand{\arraystretch}{1.3}
\begin{table}[t!]
    \centering
    \begin{tabular}{ccc} 
    \hline
	Current            & Hadronic           & Leptonic \\
    \hline
	Scalar        & $H_S=g_S~\overline{{}^3\text{He}}~{}^3\text{H}$ & 
        $j^\pm_S=\overline{e}(1\pm\gamma^5)\nu_e$ \\
		Pseudoscalar & $H_P=g_P~\overline{{}^3\text{He}}~\gamma^5~{}^3\text{H}$ & $j^\pm_P=\overline{e}~(1\pm\gamma^5)\nu_e$ \\
		Vector/Axial         & $H^\mu_{V\pm A}=\overline{{}^3\text{He}}\gamma^\mu(g_V \pm g_A \gamma^5)~{}^3\text{H}$ & $j^\mu_{V\pm A}=\overline{e}\gamma^\mu(1\pm\gamma^5)\nu_e$ \\
		Left-Tensor &  $H^{\mu\nu}_T=g_T~\overline{{}^3\text{He}}~\sigma^{\mu\nu}(1-\gamma^5)~{}^3\text{H}$ & $j^{\mu\nu}_T=\overline{e}~\sigma^{\mu\nu}(1-\gamma^5)\nu_e$ \\
		Right-Tensor &  $\tilde{H}^{\mu\nu}_T=g_T~\overline{{}^3\text{He}}~\sigma^{\mu\nu}(1+\gamma^5)~{}^3\text{H}$ & $\tilde{j}^{\mu\nu}_T=\overline{e}~\sigma^{\mu\nu}(1+\gamma^5)\nu_e$ \\
		\hline
	\end{tabular}
	\caption{Hadronic and leptonic currents expanded in the basis of gamma matrices with $\sigma^{\mu\nu}=\frac{i}{2}[\gamma^{\mu},\gamma^{\nu}]$.}
	\label{table:Currents}
\end{table}
We are here working within the regime of effective operators with interactions taking place much below the electroweak scale. From the SM weak interaction, we have taken the effective, 4-point Fermi interaction as the basis for the derivation of the decay rate. However, the vector minus axial (or $V-A$) coupling of this interaction is not the only Lorentz-invariant operator. We can also consider the presence of additional currents; these either being of the right-handed $V+A$ form or of the scalar, pseudoscalar or tensor types. By considering a basis expansion of the gamma matrices we can consider all possible contributions to the 4-point interaction \cite{Ludl:2016ane}. We break down the Lagrangian as
\begin{align}
    \label{eq:FullLag}
	\mathcal{L} = \mathcal{L}_\text{SM} + \mathcal{L}_\text{exotic} + \mathcal{L}^N_\text{exotic}, 
\end{align}
where 
\begin{align}
	\label{eq:ExL}
	\mathcal{L}_\text{exotic} = -\frac{G_F}{\sqrt{2}}V_{ud}
	\bigg(&\tilde{\epsilon}_LH^{\mu}_{V-A}j_{\mu,V+A}+\epsilon_RH^{\mu}_{V+A}j_{\mu,V-A}+\tilde{\epsilon}_RH^{\mu}_{V+A}j_{\mu,V+A} \nonumber \\
	&+\epsilon_SH_Sj^-_S+\tilde{\epsilon}_SH_Sj^+_S-\epsilon_PH_Pj^-_P-\tilde{\epsilon}_PH_Pj^+_P \nonumber\\
	&+\epsilon_TH^{\mu\nu}_Tj_{T,\mu\nu}+\tilde{\epsilon}_T\tilde{H}^{\mu\nu}_T\tilde{j}_{T,\mu\nu}\bigg),
\end{align}
gives all possible contributions involving the hadronic ($H$) and leptonic currents ($j$) as given in Table~\ref{table:Currents}. The dimensionless constants $\epsilon_i$, $\tilde\epsilon_i$ give the interaction strength relative to SM $V-A$ Fermi coupling $G_F$. The subscript denotes the hadronic part and the tilde denotes a right-handed leptonic part. An additional exotic contribution to the $V-A$ current is also possible, as represented by $\delta_\beta$ in Eq.~\eqref{eq:Lag}) but this merely leads to a re-scaling of $G_F$ so we have chosen to omit it. The $g_X$ in the hadronic currents are form factors which will, generically, depend upon the momentum exchange $Q^2$. However, the dependence can be well approximated in dipole form as $g_X(Q^2) = g_X(0)/(1-(Q^2/M_X^2))^2$ where $M_X \sim 1$~GeV $\gg Q \sim 20$~keV and thus the momentum dependence can be neglected \cite{Simkovic:2007yi}.

\subsection{Sterile Neutrinos}
We also consider contributions to tritium decay involving the emission of a keV-scale sterile neutrino $N$. They can be defined analogously to the above with the substitution of leptonic currents $j\to J$ that now involve the sterile state $\nu_e\to N$,
\begin{align}
    \label{eq:ExLN}
	\mathcal{L}^N_\text{exotic} = -\frac{G_F}{\sqrt{2}}V_{ud}
	\bigg(&\epsilon^N_LH^{\mu}_{V-A}J_{\mu,V-A}+\tilde{\epsilon}^N_LH^{\mu}_{V-A}J_{\mu,V+A} \nonumber\\
	&+\epsilon^N_RH^{\mu}_{V+A}J_{\mu,V-A}+\tilde{\epsilon}^N_RH^{\mu}_{V+A}J_{\mu,V+A} \nonumber\\
	&+\epsilon^N_SH_SJ^-_S+\tilde{\epsilon}^N_SH_SJ^+_S-\epsilon^N_PH_PJ^-_P-\tilde{\epsilon}^N_PH_PJ^+_P \nonumber\\
	&+\epsilon^N_TH^{\mu\nu}_TJ_{T,\mu\nu}+\tilde{\epsilon}^N_T\tilde{H}^{\mu\nu}_T\tilde{J}_{T,\mu\nu}\bigg).
\end{align}
Here, the $\epsilon^N_i$, $\tilde\epsilon^N_i$ equivalently parametrize the strength of the exotic currents. This Lagrangian also has an additional purely left-handed contribution parametrized by $\epsilon^N_L$.

In calculating the sterile contributions, we will mainly consider a simplified scenario with one massless, active electron neutrino (the SM contribution) and one sterile neutrino with mass $m_N \lesssim 18$~keV. The active and sterile states can be expressed as mixtures of the mass eigenstates, $\nu_{1,2}$, with mixing strength $0 \leq V_{eN} \ll 1$ here taken to be real and, without loss of generality, positive,
\begin{align}
\label{eq:SteMix}
	\nu_e &= \sqrt{1-V_{eN}^2} ~ \nu_1 +  V_{eN} ~ \nu_2, \nonumber\\ 
	  N     &=        -V_{eN} ~ \nu_1 + \sqrt{1-V_{eN}^2} ~ \nu_2.
\end{align}
The state $\nu_1$ can thus be seen as mostly active whereas $\nu_2$ is mostly sterile when $V_{eN}$ is small.

\subsection{Existing Constraints}
\begin{table}[t!]
    \centering
    \begin{tabular}{ccc}
    \hline
     Coupling & $|\operatorname{Re}\epsilon|$ & $|\operatorname{Im}\epsilon|$\\
     \hline
     $\epsilon_S$ & $8\times 10^{-3}$ &
     $1\times 10^{-2}$
     \\
     $\tilde{\epsilon}_S$ & $1.3\times10^{-2}$ & $1.3\times 10^{-2}$ \\
     $\epsilon_P$ & $4.6\times10^{-7}$ & $2\times 10^{-4}$\\
     $\tilde{\epsilon}_P$ & $2\times10^{-4}$ & $2\times 10^{-4}$\\
     $\epsilon_T$& $1\times10^{-3}$ & $1\times 10^{-3}$\\
     $\tilde{\epsilon}_T$&  $3\times10^{-3}$ &
     $3\times 10^{-3}$\\
     $\tilde{\epsilon}_L$& $6\times10^{-2}$ & - \\
     $\epsilon_R$  & $5\times10^{-4}$ & $5\times 10^{-4}$\\
     $\tilde{\epsilon}_R$ & $5\times10^{-3}$ &
     $5\times 10^{-3}$\\
     \hline
    \end{tabular}
    \caption{Experimental upper bounds on the real and imaginary parts of the exotic coupling strengths at $90\%$~CL. Adapted from \cite{Ludl:2016ane, Falkowski:2021} and with experimental sources given in the text.}
    \label{tab:upper-bounds}
\end{table}
Experiments have sought to measure or place upper bounds upon the strengths of exotic currents. The upper bounds quoted in Table~\ref{tab:upper-bounds} mostly come from $\beta$-decay experiments and from LHC processes, with the stronger of these bounds being given for each exotic current parameter. The current most stringent upper bounds on $\epsilon_P$ and $\tilde{\epsilon}_P$ (both the real and imaginary parts) come from low-energy experiments involving pion decay, more precisely the ratio in pion decay of $\pi\to e \nu_e$ to $\pi\to\mu\nu_\mu$ which is dependent upon the pseudoscalar current \cite{Cirigliano:2013,Falkowski:2021}. The upper limits on $\operatorname{Re}(\epsilon_{R,L})$ come from $\beta$-decay experiments through placing bounds on the unitarity of the CKM matrix \cite{cirigliano2010semileptonic}. The limit on $\operatorname{Im}(\epsilon_R)$ has been placed by the emiT collaboration by measuring spin-angle correlations in $\beta$-decay \cite{chupp2012search, mumm2011new}. The largest possible values of $\operatorname{Re}(\epsilon_{S,T})$ compatible with the excluded regions of their respective parameter spaces is of the same order of magnitude for both the low-energy and LHC probes, where the low-energy probe involves $0^+\to 0^+$ nuclear $\beta$-decays \cite{hardy2009superallowed} and radiative pion decay from the PIBETA collaboration \cite{bychkov2009new}, for the scalar and tensor currents, respectively. These bounds are projected to reach a precision level of $0.1\%$ through measurements of neutron \cite{dubbers2008clean, alarcon2007precise, wilburn2009measurement, povcanic2009nab, markisch2009new} and $\operatorname{^{6}{He}}$ \cite{knecht2011high} decays. The $\beta$-decay and LHC bounds are also similar in the case of $\operatorname{Im}(\epsilon_T)$, where the low-energy limit is based on spin-angle correlations in $\operatorname{^{8}{He}}$. In the cases of $\operatorname{Im}(\epsilon_S)$ and $\tilde{\epsilon}_{R,S,T}$ (both the real and imaginary parts) the strongest bounds come from LHC measurements. In order for the low-energy probes to match this sensitivity, the measurement of neutrino angular correlations in Gamow-Teller transitions, $a_{GT}$, needs to be improved to the level of $\delta a_{GT}/a_{GT}\sim 0.05\%$ \cite{Cirigliano:2013}. If sterile neutrinos are Majorana fermions, limits on the operator coefficients can also be derived from neutrinoless double $\beta$ decay experiments \cite{Tong:2021}. 

\section{Tritium Decay} 
\label{Calculation}

\subsection{Differential Decay Rate}

Two different $\beta$-decay distributions will be of interest to us: the energy spectrum as a function of the electron energy and the angular spectrum as a function of the angle between the electron's momentum and the tritium's spin. Both quantities can be calculated from the fully differential decay rate,
\begin{align}
	d\Gamma=\frac{1}{2m_\text{H}} 
	\frac{d^3\boldsymbol{p}_e}{(2\pi)^3 2E_e}
	\frac{d^3\boldsymbol{p}_\nu}{(2\pi)^3 2E_\nu}
	\frac{d^3 \boldsymbol{p}_\text{He}}{{(2\pi)}^3 2E_\text{He}} |T|^2 
	(2\pi)^4 \delta^4 (P_\text{H} - P_\text{He} - P_e - P_\nu).
\end{align}
where the lower case $\boldsymbol{p_i}$ denote the 3-momenta of the associated particles, whereas $P_i$ are the 4-momenta, and $|T|^2$ is the squared matrix element of the process.

Evaluating in the tritium rest frame, $P_\text{H} = (m_\text{H}, \mathbf{0})$, using the energy-momentum conserving delta function and expressing in spherical coordinates for the electron and neutrino momenta gives
\begin{align}
	\frac{d\Gamma}{dE_ed\Omega_e} = 
	\frac{C(E_e)}{2^9\pi^5 m_\text{H}}\int_\Omega d\Omega_\nu \int_{E_\nu^-}^{E_\nu^+} dE_\nu |T|^2
	\delta\left(\cos{\theta_{e\nu}} 
	- \frac{{\boldsymbol{p}_\text{He}^2}
	- {\boldsymbol{p}_e^2} 
	- {\boldsymbol{p}_\nu^2}}{2|\boldsymbol{p}_e||\boldsymbol{p}_\nu|}\right).
\label{eq:AngInt}
\end{align}
Here, $\Omega_i$ is the solid angle of emission of the electron or neutrino (which we choose to be oriented such that the polar angle is with respect to the spin direction of the tritium nucleus) and $\theta_{e\nu}$ is the angle between the electron and neutrino three-momenta. The factor $C(E_e)$ collects the theoretical corrections as described in Sec.~\ref{sec:Corrections}. In the integral over the neutrino energy, $E^{\pm}_\nu(E_e)$ are the upper and lower bounds on the neutrino energy for a given electron energy~\cite{Ludl:2016ane},
\begin{align}
	E_\nu^{\pm}(E_e) = 
	\frac{(m_\text{H} - E_e)(m_\text{H} y + m_\nu^2 + m_\text{He} m_\nu) 
	\pm |\boldsymbol{p}_e|\sqrt{(m_\text{H}y(m_\text{H}y + 2m_\text{He}m_\nu)}}{m_{12}^2}.
\end{align}
Here, we define
\begin{align}
\label{eq:m12sq}
	m_{12}^2 = (p_\text{H} - p_e)^2 = m_\text{H}^2 - 2m_\text{H}E_e + m_e^2,
\end{align}
and 
\begin{align}
\label{eq:y}
	y = E_e^\text{max} - E_e,
\end{align}
where $E_e^\text{max}$ is the maximum electron energy for tritium decay with $E_e^\text{max} - m_e \approx 18.59$~keV \cite{AUDI2003337, Masood:2007rc}. It is worth noting that $E_e^{max}$ and hence $y$ have an implicit dependence on the neutrino mass,
\begin{equation}
	E_e^{\rm{max}} = \frac{m_\text{H}^2 + m_e^2 - (m_\text{He} + m_\nu)^2}{2m_\text{H}}.
\end{equation}

In general, the particles will be in particular spin states. Typically these spins are summed or averaged to derive the decay rate corresponding to the source having an equal probability of being spin up or down. If we wish to consider the impact of the spin direction on the emission of the electron (such as when the ensemble is polarised) then it is more useful to preserve the choice of spin state. One way to achieve this is by projecting onto the required spin state whilst still summing over the spins. This is done with the projection operator $P_s = \frac{1}{2}(1 + \gamma^5 \slashed{S})$ \cite{Pal2014} with $S^\mu = (\boldsymbol{p}\cdot\hat{\boldsymbol{s}}/m, \hat{\boldsymbol{s}} + (\boldsymbol{p}\cdot\hat{\boldsymbol{s}})/(m(E + m)))$ where the mass, spin and momentum are for the relevant particle as evaluated in a given frame. In this case the projector will be applied to the tritium which in its rest frame gives $S^\mu = (0,\hat{\boldsymbol{s}})$~\footnote{The projector $P_s$ is chosen to satisfy $P_s u_{\boldsymbol{p}, s'} = \delta_{s,s'} u_{\boldsymbol{p},s'}$ and $P_s v_{\boldsymbol{p}, s'} = \delta_{s,s'} v_{\boldsymbol{p}, s'}$ where $s, s' = \downarrow,\uparrow$ and $u,v$ are the spinors of the associated particle.}. 

Generally, the tritium ensemble can be partially polarised with a polarisation factor of 
\begin{align}
	f = \frac{N^{\uparrow} - N^{\downarrow}}{N^{\uparrow} + N^{\downarrow}},
\end{align}
where $N^{\uparrow(\downarrow)}$ denotes the number of spin up (down) tritium nuclei or, alternatively, the population probabilities. This arises because when we are considering the net impact of polarisation across the ensemble we must sum over all the up and down states with $N^{\uparrow}\to N^{\downarrow}$ corresponding to $\hat{s} \to -\hat{s} \Rightarrow S \to -S$ and thus there is a cancellation between them. A polarisation of the ensemble will lead to an angular non-isotropy. When considering the angular distribution we will choose the maximal value $f = 1$, i.e., total polarisation, but any value can be easily used by taking $S\to fS$ in the matrix element or $\hat{s} \to f\hat{s}$ in the decay rate. No polarisation corresponds to $f = 0$.

When polarised, a correlation arises between the electron's momentum direction and the spin of the tritium. As the spin dependence is always linear (due to the linearity of the singly included projector), the decay rate can be written as a linear sum of spin-independent and -dependent terms. Thus in general the double differential distribution with respect to the electron energy $E_e$ and its solid angle takes the form
\begin{align}
	\frac{d\Gamma}{dE_e d\Omega_e} =a(E_e)+b(E_e)\cos\theta_e, 
\label{eq:Pol}
\end{align}
where $a(E_e)$, $b(E_e)$ are functions which will depend on the choice of Lagrangian and $\theta_e$ is the angle between the electron momentum and the tritium spin. 

Because of the $\cos\theta_e$ dependence, when integrating over all electron angles, the contribution from $b(E_e)$ will vanish leaving the energy distribution,
\begin{align}
\label{eq:dGammadE-general}
	\frac{d\Gamma}{dE_e} = 4\pi a(E_e).
\end{align}
The total tritium decay rate is then
\begin{align}
    \label{eq:Gamma}
	\Gamma = 4 \pi \int_{m_e}^{E_e^\text{max}} a(E_e) dE_e,
\end{align}
Alternatively, the double differential distribution can be integrated over the electron energy to give the angular distribution
\begin{align}
	\frac{d\Gamma}{d\cos\theta_e} = 
	\frac{\Gamma}{2}(1+k\,\cos\theta_e),
\label{eq:IntPol}
\end{align}
with the angular correlation factor
\begin{align}
    k=\int_{m_e}^{E_e^{max}}b(E_e)dE_e\Bigg/\int_{m_e}^{E_e^{max}}a(E_e)dE_e.
\end{align}

\subsection{Theoretical Correction Factors}
\label{sec:Corrections}
The factor $C(E_e)$ in Eq.~\eqref{eq:AngInt} denotes additional theoretical correction factors to the spectrum which appear as multiplicative overall factors. See Appendix~\ref{App:Cor} for full expressions and further details on all theoretical corrections. The most significant of these is the Fermi factor, $F(E_e)$ which corrects for the electromagnetic interaction between the emitted electron and the final state nucleus. All of the other multiplicative correction factors are close to unity for the vast majority of the spectrum. For atomic tritium, the relevant corrections are: radiative corrections to the spectrum, $G(E_e)$; the nuclear screening from the orbital electron, $S(E_e)$; finite size nucleus effects, $L(E_e)B(E_e)$ and recoil corrections to the nuclear Coulomb field, $Q(E_e)$. Thus the overall factor is
\begin{align}
    C(E_e) = F(E_e)\times G(E_e)\times S(E_e)\times L(E_e)\times B(E_e)\times Q(E_e).
\end{align}
\begin{figure}
    \centering
    \includegraphics[width=0.6\textwidth]{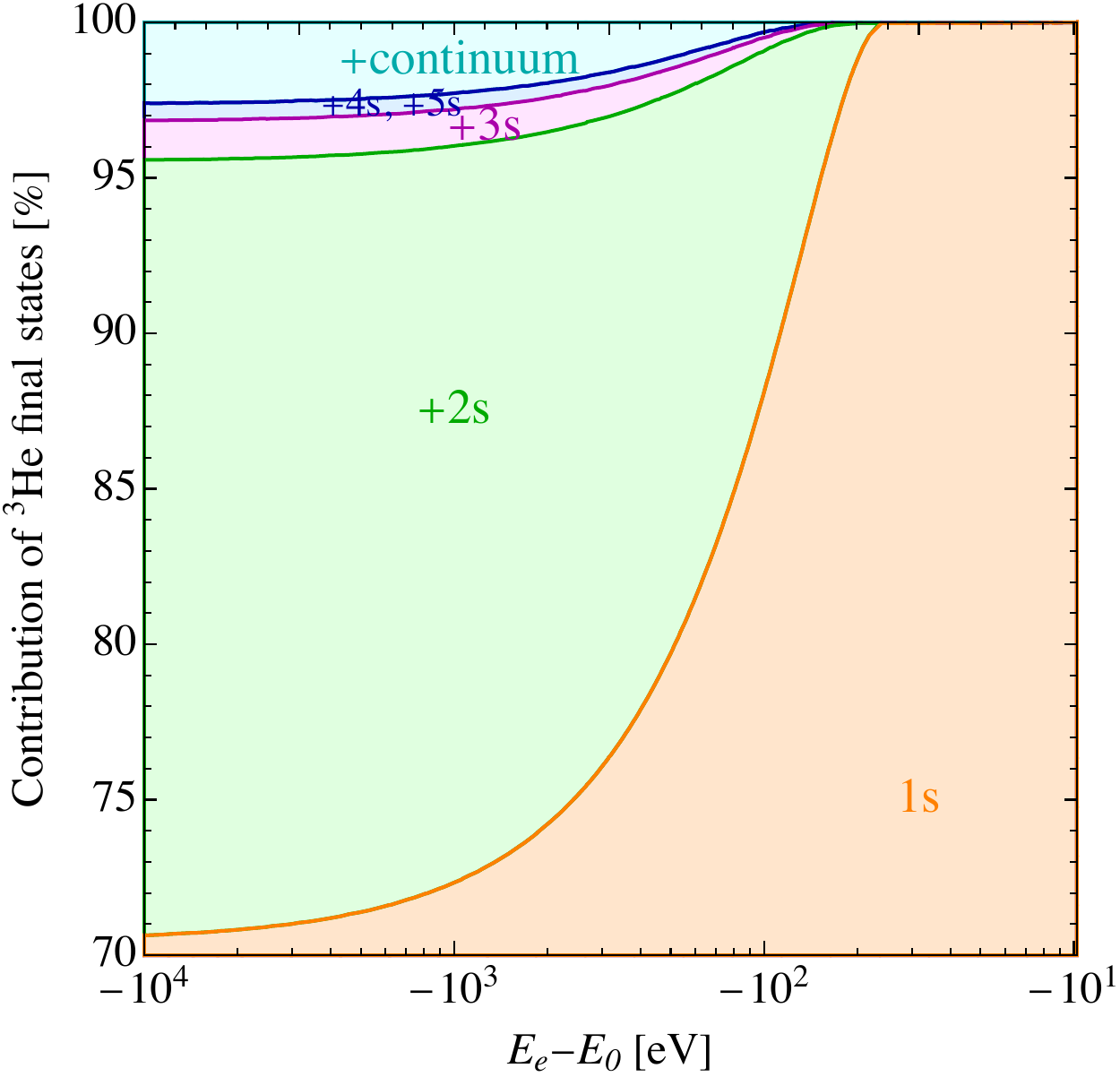}
    \caption{Fraction of decays to helium energy levels $1s$ (ground state), $2s$, $3s$, $4s + 5s$ of the orbital electron, as well emission of the orbital electron (continuum) as a function of the $\beta$-decay electron energy below the endpoint.}
    \label{fig:EnLev}
\end{figure}
The other significant effect is that of the quantum mechanical interactions between the emitted and orbital electron. Whilst the orbital electron is initially in a 1s state, it is possible for it to end up in an excited state or even be emitted into the continuum. This leads to a spectrum of endpoints as shifted by the helium energy levels. The total decay rate is thus a sum over the decay rates to different orbital states or the continuum, as shown in Eq.~\eqref{eq:OrbitSum}. The probability of decay to each state as a function of $\beta$-electron energy is shown in Fig.~\ref{fig:EnLev} with the endpoint shifts and asymptotic probabilities given in Table \ref{table:emission}. The significance of having multiple endpoints is that instead of having a single kink in the energy spectrum for where each mass eigenstate becomes kinematically allowed, there are now a series of kinks of a fixed separation and sharpness. For this paper, we have assumed an energy resolution of around 100~eV which is larger than the roughly 50~eV separation between energy levels. However, a future sterile neutrino experiment with a higher resolution could use the multi-kink signature as a specific signal to increase the search power.
\begin{table}[t!]
	\centering
	\begin{tabular}{ccr} 
		\hline
		$\text{He}^+$ State & Energy [eV] & $|T_{fi}^{(0)}|^2$ \\ 
		\hline
		$1s$      & 0\phantom{,0}     & 70.36\% \\ 
		$2s$      & 40.81 & 24.98\% \\
		$3s$      & 48.37 &  1.27\% \\
		$4s$      & 51.02 &  0.38\% \\
		$5s$      & 52.24 &  0.17\% \\
		continuum & $\geq$54.42 &  2.63\% \\ 
		\hline
	\end{tabular}
	\caption{Excitation energy above the ground state associated with final helium orbital electron states and the corresponding asymptotic branching ratios for large $\beta$-electron energy.}
	\label{table:emission}
\end{table}

\subsection{Standard Model Case}
Following the above general discussion on the tritium decay distribution, we first calculate the SM result for the energy and angular distributions following the procedure above and using the Lagrangian given by Eq.~\eqref{eq:Lag}. The matrix element for this contribution is given by 
\begin{align}
	|T_\text{SM}|^2 &= 
	16 G_F^2 |V_{ud}|^2 \nonumber\\
	&\times\left\{
	(g_A+g_V)^2 (P_e\cdot P_\text{He}) (P_{\nu} \cdot P_\text{H}) 
	 +(g_A-g_V)^2(P_e\cdot P_\text{H}) (P_{\nu} \cdot P_\text{He}) \right.\nonumber\\
	&\quad +(g_A^2-g_V^2)m_\text{H}m_\text{He}(P_e \cdot P_{\nu}) \nonumber\\
	&\quad +(g_A^2-g_V^2)m_\text{He}\left[(P_\text{H}\cdot P_{\nu}) (P_e\cdot S) 
	 - (P_\text{H}\cdot P_e) (P_{\nu}\cdot S)\right] \nonumber\\
	&\quad \left.+(g_A-g_V)^2 m_\text{H} (P_{\nu} \cdot P_\text{He})(P_e \cdot S) 
	 -(g_A+g_V)^2 m_\text{H} (P_e \cdot P_\text{He}) (P_{\nu} \cdot S)\right\}.
\label{eq:Mat}
\end{align}
Considering the case where the tritium ensemble is unpolarized ($f = 0$) and using Eq.~\eqref{eq:AngInt} gives the energy distribution
\begin{align}
	4\pi a_{\text{SM}}(E_e) &= 
	\frac{G_F^2 |V_{ud}|^2}{2\pi^3} C(E_e) \sum_{i=1}^3 |U_{ei}|^2 \frac{m_\text{H}^2|\boldsymbol{p}_e|}{m_{12}^2} \widetilde{y_i} \Theta(y_i)
	 \nonumber\\ 
	&\times\left\{(g_V\!+\!g_A)^2 \Bigg[\frac{m_\text{H}(m_\text{H}-E_e)}{m_{12}^2}\frac{m_\text{H}E_e-m_e^2}{m_{12}^2}
	\left(y_i\!+\!\mu_i m_{\nu,i}\right)
	\left(y_i\!+\!\mu_i m_\text{He}\right)\right. \nonumber\\
	&\quad -\frac{m^2_\text{H}|\boldsymbol{p}_e|^2}{3m_{12}^4} \widetilde{y_i}^2\Bigg] +(g_V\!-\!g_A)^2E_e\left(y_i+m_{\nu,i}\frac{m_\text{He}}{m_\text{H}}\right) \nonumber \\
	     &\quad  +(g_A^2\!-\!g_V^2)m_\text{He}\frac{m_\text{H}E_e-m_e^2}{m_{12}^2}\left(y_i+\mu_i m_{\nu,i}\right)  \Biggr\}, 
\label{eq:FulBet}
\end{align}
where we sum over all the three mass eigenstates of active neutrinos with the Heaviside function $\Theta(y_i)$ ensuring energy conservation of each individual contribution. The quantities $m_{12}^2$ and $y_i$ are defined in Eq.~\eqref{eq:m12sq} and Eq.~\eqref{eq:y}, respectively, $\mu_i = (m_{\nu,i} + m_\text{He})/m_\text{H}$ and
\begin{align}
    \widetilde{y_i} = 
    \sqrt{y_i\left(y_i + m_{\nu,i}\frac{2m_\text{He}}{m_\text{H}}\right)}.
\end{align}
The expression in Eq.~\eqref{eq:FulBet} matches \cite{Simkovic:2007yi} with minor typos corrected. It can be simplified using the approximations $m_\text{H} \approx m_\text{He} \gg E_e, m_e, m_\nu$ to give
\begin{align}
\label{eq:AppBet}
	4\pi a_{\text{SM}}(E_e) &\approx 
	\frac{G_F^2|V_{ud}|^2}{2\pi^3} C(E_e) (g_V^2+3g_A^2) m_\text{H} m_\text{He} \frac{m_\text{He}|\boldsymbol{p}_e|}{m_{12}^2}\frac{m_\text{H}E_e-m_e^2}{m_{12}^2} \nonumber\\
	&\times \sum_{i=1}^3 |U_{ei}|^2 \Theta(y_i) \sqrt{y_i\left(y_i+m_{\nu,i}\frac{2m_\text{He}}{m_\text{H}}\right)}
	\left(y_i+m_{\nu,i}\frac{m_{\nu,i}+m_\text{He}}{m_\text{H}}\right).
\end{align}
This is usually expressed in the familiar form \cite{shrock1980new, Huang:2019tdh}
\begin{align}
\label{eq:dGammadE-mbeta}
	4\pi a_{\text{SM}}(E_e)\approx
	\frac{G_F^2|V_{ud}|^2}{2\pi^3} C(E_e) 
	(g_V^2+3g_A^2)m_\text{H}m_\text{He}
	\frac{m_\text{H}|\boldsymbol{p}_e|}{m_{12}^2}\frac{m_\text{H}E_e-m_e^2}{m_{12}^2}  
	y_0\sqrt{y_0^2-m_\beta^2},
\end{align}
using the effective single $\beta$-decay mass $m_\beta^2 = \sum_{i=1}^3|U_{ei}|^2 m_i^2$, and where $y_0 = y(m_\nu = 0)$ ensures that all of the neutrino mass dependence has been combined into $m_\beta$.

\begin{figure}[t!]
	\centering
	\includegraphics[width=0.75\textwidth]{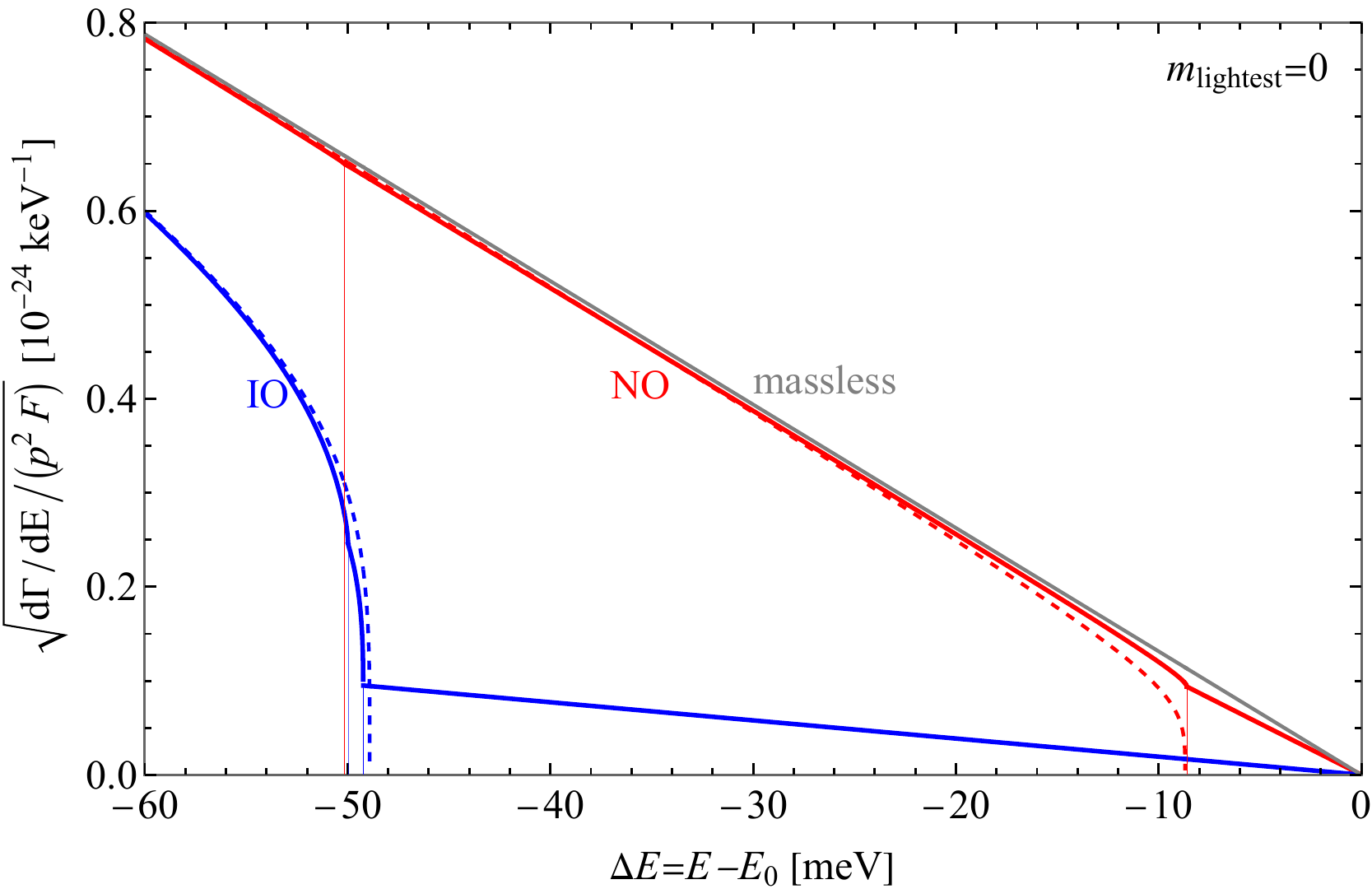}
	\caption{Kurie plot of the tritium energy distribution as measured from the zero-mass endpoint for a massless lightest neutrino with NO (red) and IO (blue) of the three active neutrinos. The red (blue) dashed lines correspond to the effective $m_\beta$ approximation for NO (IO) as given by Eq.~\eqref{eq:mbeta}. The gray line corresponds to the single massless active neutrino case.}
    \label{fig:Kurie}
\end{figure}
Using the full expression in Eq.~\eqref{eq:FulBet} with three active neutrinos will result in three distinct endpoints. Instead, the simplified expression in Eq.~\eqref{eq:dGammadE-mbeta} has a single endpoint at $E_e = E_0 - m_\beta$. The difference is illustrated in Fig.~\ref{fig:Kurie}, showing the Kurie plot of the energy distribution near the endpoint using the exact expression and the approximation using $m_\beta$, for a massless lightest neutrino in both normally and inversely ordered scenarios. As described in \cite{Huang:2019tdh}, the difference in the spectrum when using this approximated version is negligible for energy resolutions much larger than the neutrino mass or when looking at points far from the endpoint. For our calculations we are looking across the whole spectrum with an energy resolution much larger than the neutrino mass and thus this difference will be negligible.

The angular correlation term in the SM case is given by
\begin{align}
	b_\text{SM}(E_e)&=
	-\frac{G_F^2|V_{ud}|^2}{8\pi^4} C(E_e)\frac{m_\text{H}}{m_{12}^2}|\boldsymbol{p}_e|^2
	 \widetilde{y_i} \nonumber\\
	&\times\left\{\left[(g_A-g_V)^2 m_\text{H} 
	+ (g_A^2-g_V^2)m_\text{He}\frac{m_\text{H}(m_\text{H}-E_e)}{m_{12}^2}\right.\right. \nonumber\\
	&\quad\,\,\,+(g_A^2-g_V^2)\frac{m_\text{H}m_\text{He}}{m_{12}^2}E_e 
	 +(g_A+g_V)^2\frac{m_\text{H}}{m_{12}^2}(\alpha-m_e^2)\nonumber\\ 
	&\quad\,\,\,\left.-(g_A+g_V)^2\frac{m_\text{H}^2}{m_{12}^2}\left(y_i+\mu_i m_{\nu,i}\right)
	  \frac{m_\text{H}(m_\text{H}-E_e)}{m_{12}^2}\right]
	  \left(y_i+\mu_i m_{\nu,i}\right)\nonumber\\
	&\quad\,\,\,\left.-(g_A-g_V)^2m_{\nu,i}^2 
	-\frac{1}{3}(g_A+g_V)^2\frac{m_\text{H}^3(m_\text{H}-E_e)}{m_{12}^4}\widetilde{y_i}^2\right\}. 
\label{eq:FulAng}
\end{align}
where $\alpha=m_{\text{H}}E_e^{max}+m_{\nu_i}^2+m_{\text{He}}m_{\nu_i}$. This expression can be greatly simplified under the same approximation, as before, $m_\text{H}\approx m_\text{He} \gg E_e,m_e,m_{\nu_i}$,
\begin{align}
    \label{eq:SMAng}
	b_\text{SM}(E_e)\approx 
	-\frac{2g_A^2 - 2g_Ag_V}{g_V^2 + 3g_A^2}
	\frac{|\boldsymbol{p}_e|}{E_e} a_{\text{SM}}(E_e) \approx -0.12\frac{|\boldsymbol{p}_e|}{E_e} a_{\text{SM}}(E_e).
\end{align}
Using Eqs.~\eqref{eq:Gamma}, \eqref{eq:FulBet} and \eqref{eq:SMAng} gives a total half-life of $T_{1/2}=\ln(2)/\Gamma=12.6$~yr and an angular correlation factor of $k_\text{SM} = -0.0154$ in the SM. The total decay rate $\Gamma$ and the angular correlation factor $k$ can also be obtained from experimental measurements. The decay rate can be calculated from the half-life by $\Gamma = \ln(2)/T_{1/2}$. The angular correlation can be found by measuring the proportion of aligned, i.e., in the same hemisphere as the tritium spin, to anti-aligned, i.e., in the opposite hemisphere electrons which takes the ratio $(1+k/2)~:~(1-k/2)$.
 
This angular effect occurs due to the electron's helicity and chirality. The emitted electron and anti-neutrino can be either in an anti-aligned spin state ($S=0$), corresponding to the vector hadronic current, or an aligned spin state ($S=1$), corresponding to the axial hadronic current \cite{Pal2014}. The $S=0$ emission is isotropic in spin space. However, the $S=1$ state requires the spins to not sum within the opposite hemisphere to the tritium's polarisation in order to conserve angular momentum. This preferentially places the electron's spin within the same hemisphere as the polarisation. As the electron is left-chiral (and thus preferentially left-helical) it is more likely to be emitted with momentum opposite to the spin and thus in the opposite hemisphere. Thus the decay rate to an electron parallel to the tritium's polarisation is reduced. Crucially, this effect is proportional to $\hat{\boldsymbol{p}}_e\cdot\hat{\boldsymbol{s}}$ and thus once integrated over the electron's solid angle it vanishes. This is to be expected as polarising the tritium merely orientates it rather than changing the internal dynamics that lead to decay. Such correlations could also be calculated for the momentum of the neutrino or the helicity of the electron. Whilst measuring the electron's helicity could also be informative, designing an experiment to simultaneously obverse this and the spectrum is prohibitively challenging so we will not consider it further here. 

\subsection{Exotic Currents}

In addition to the SM contribution we include the effects of the exotic currents in Eq.~\eqref{eq:ExL}. This will lead to additional contributions to the energy spectrum and the angular distribution, when polarised. When squaring the matrix element we will get SM, purely exotic and SM-exotic interference contributions to the decay rate,
\begin{align}
	a(E_e) &=  a_\text{SM}(E_e) + \text{Re}(\epsilon_Y) a_{LL,Y}(E_e) 
	+ |\epsilon_Y|^2 a_Y(E_e), \\
	b(E_e) &=  b_\text{SM}(E_e) + \text{Re}(\epsilon_Y) b_{LL,Y}(E_e) 
	+ |\epsilon_Y|^2 b_Y(E_e).
\end{align}
For all of the currents with an exotic left-handed leptonic current, the interference term will dominate as it is linear in $\epsilon \ll 1$ with the potential for cancellation between the linear and quadratic terms when $\epsilon$ is larger. Conversely, for the right-handed leptonic currents, i.e. those parametrised by $\tilde{\epsilon}$, a chiral flip is required and thus the interference term is proportional to the neutrino and electron masses. These interference terms are thus very small and the quadratic terms will dominate. 

\begin{figure}[t!]
    \centering
    \includegraphics[width=0.47\textwidth]{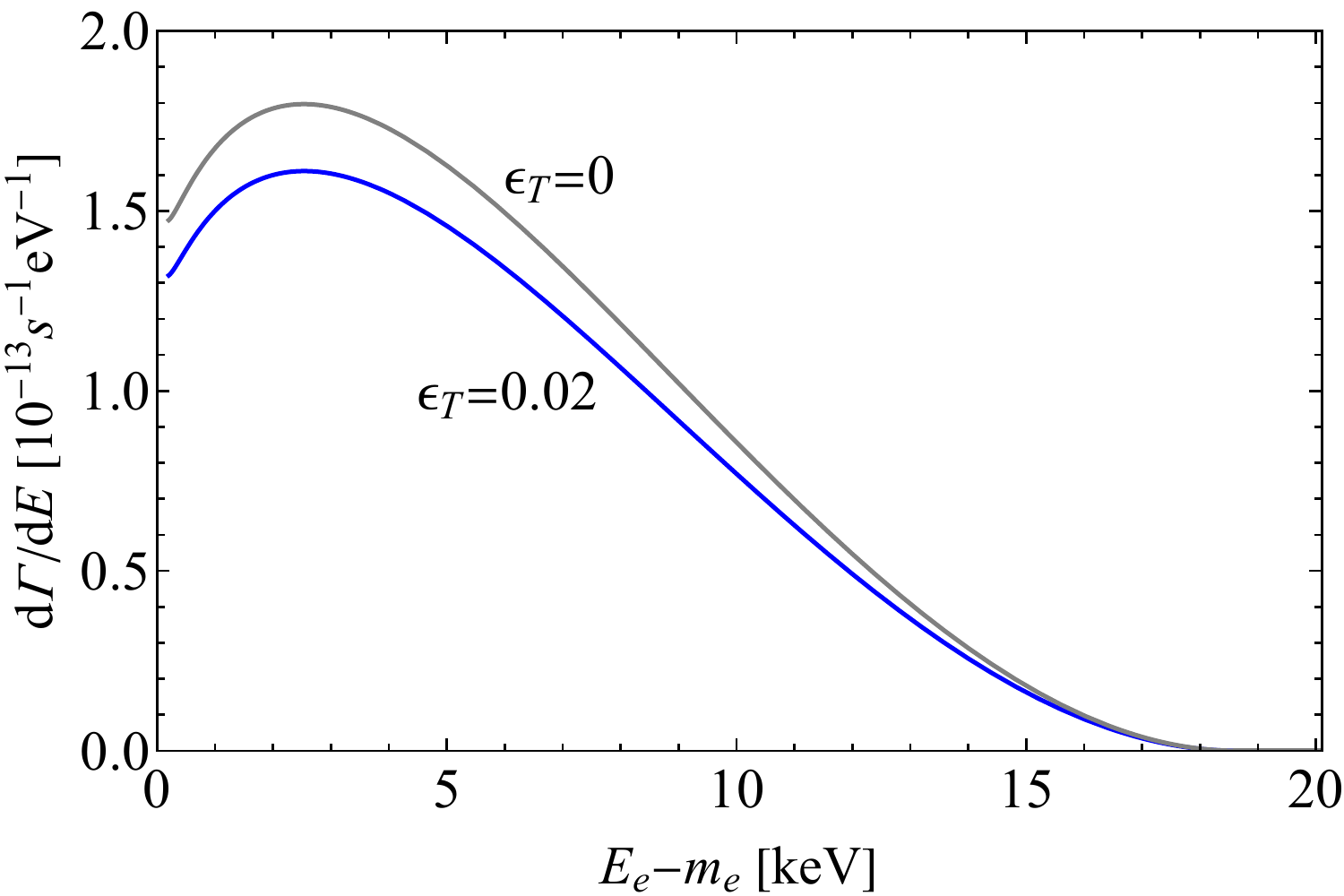}
    \includegraphics[width=0.52\textwidth]{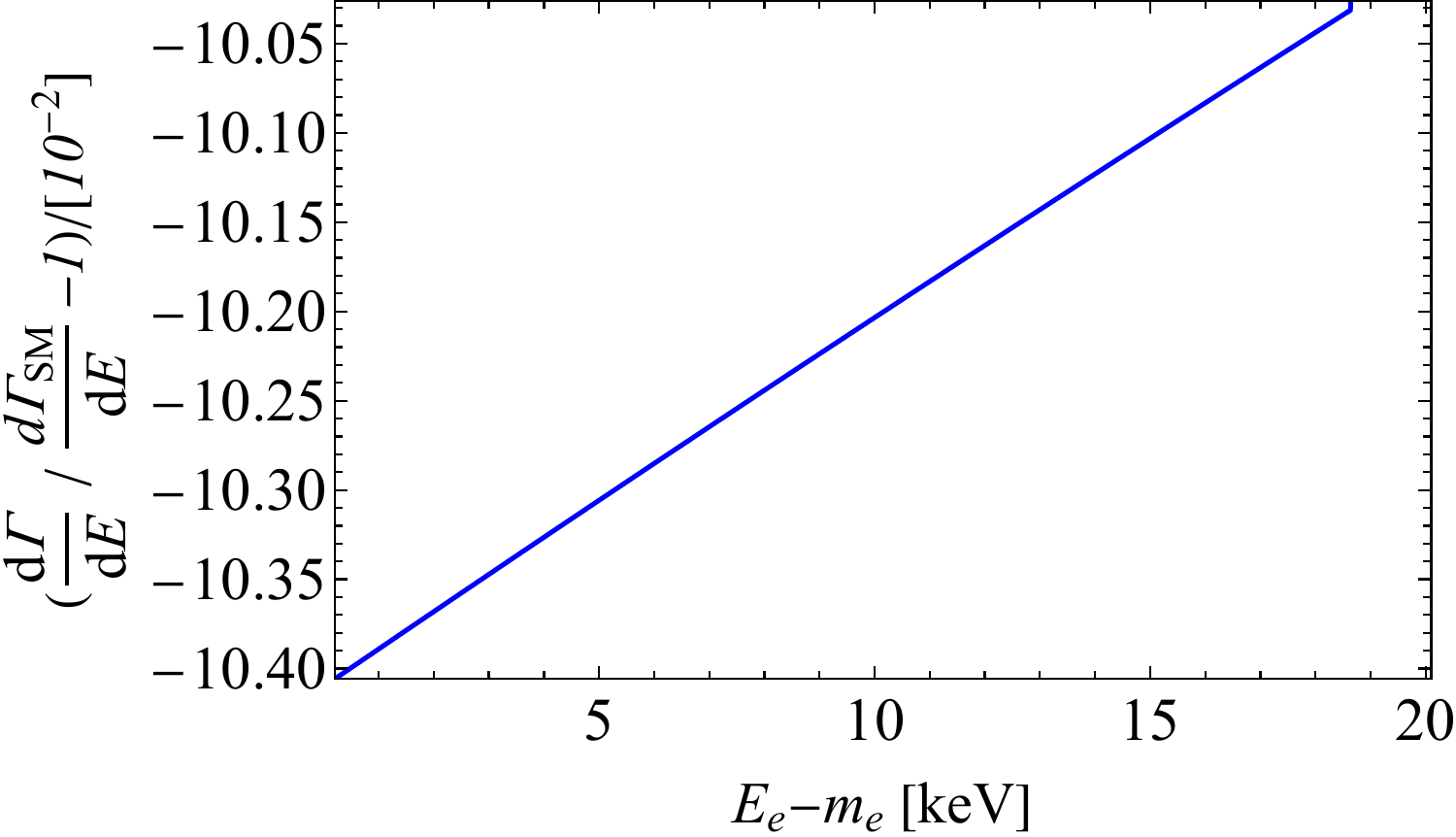}
    \caption{Left: Electron energy distribution, in terms of the electron kinetic energy, in the SM ($\epsilon_T = 0$) and with a left-handed tensor current, $\epsilon_T = 0.02$. Right: Relative deviation between the two spectra.}
	\label{fig:tensor}
\end{figure}
We use the form factor values $g_S=1.02$, $g_P=349$, $g_T=1.02$ from \cite{PhysRevLett.112.042501, PhysRevLett.115.212002} which are calculated from phenomenological studies and Lattice QCD. The impact of an additional exotic current can be seen in Fig.~\ref{fig:tensor}, where the presence of a tensor current reduces the decay rate nearly homogeneously across the energy spectrum. See Appendix \ref{App:full-expressions} for full expressions for all exotic currents considered, where we give both the individual exotic contribution as well as the interference with the SM.

\subsection{Sterile Neutrinos}
If we consider the sterile state as only appearing through its mixing with the active neutrinos, as given by Eq.~\eqref{eq:SteMix}, then its additional contribution to the decay rate and angular distribution is simple. The decay rate expressions are the same as in Eqs.~\eqref{eq:FulBet},~\eqref{eq:FulAng} but with the sum over neutrino masses now going from $i=1,\dots,3+n_s$ with $n_s$ sterile states and corresponding masses $m_4,\dots, m_{n_s+3}$ and mixing matrix elements $V_{e4},\dots,V_{e{(n_s+3)}}$. 

\begin{figure}[t!]
	\centering
	\includegraphics[width=0.75\textwidth]{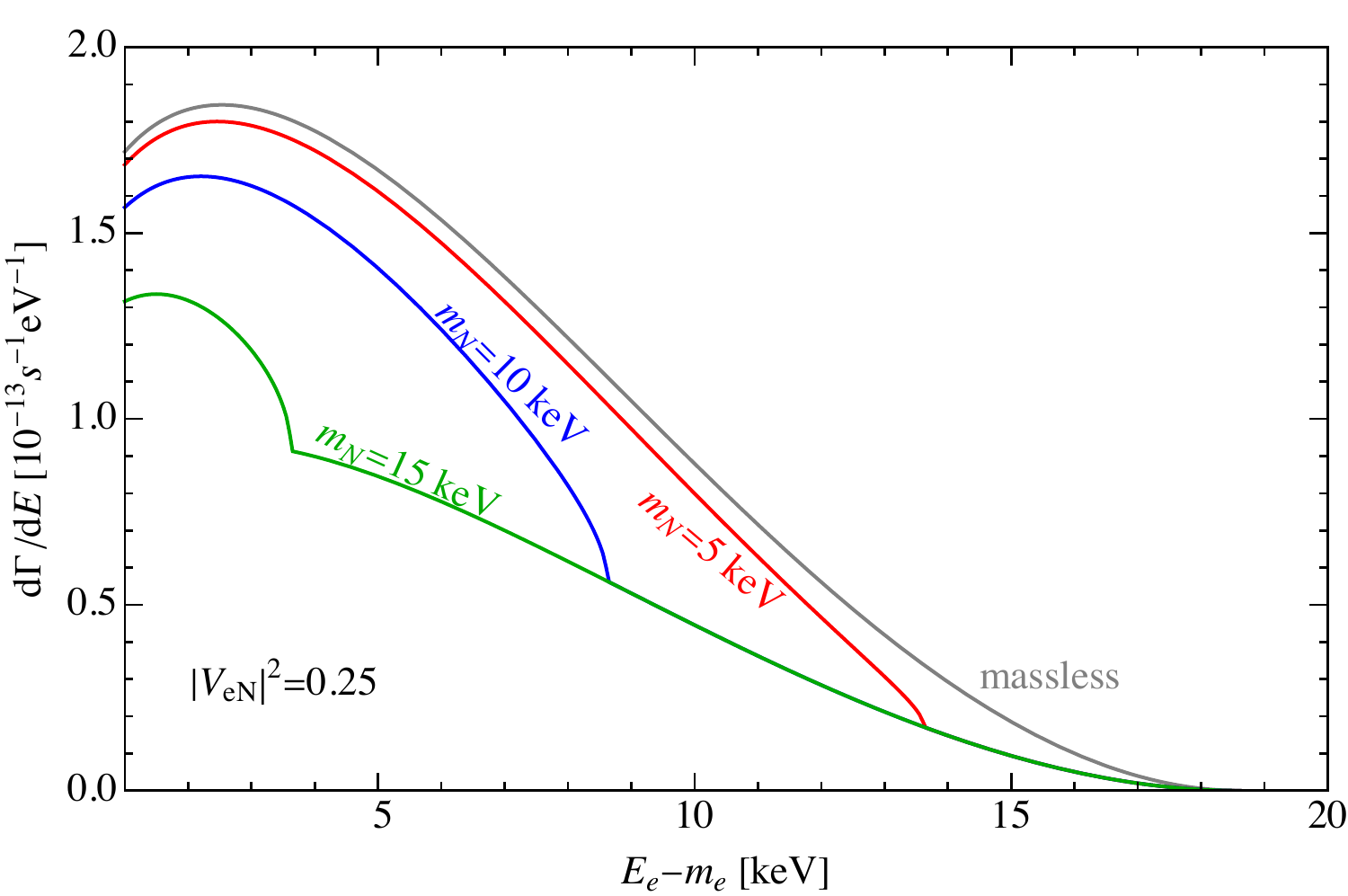}
	\caption{Differential decay rate in terms of the electron kinetic energy, for a massless and various heavy, mostly-sterile neutrinos with a mixing of $|V_{eN}|^2 = 0.25$.}
	\label{fig:spectrum}
\end{figure}
The impact of a sterile neutrino on the energy distribution in such a case can be seen in Fig.~\ref{fig:spectrum}. Within the energy spectrum, the distinctive kink from the neutrino mass is visible at an energy $E^{\rm{max}}_e-m_{N}$. The spectrum for energies below the kink is reduced as the heavier sterile state reduces the phase space with heavier masses giving a greater reduction. The spectrum for energies above the kink is also reduced but is the same for all sterile masses as it consists only of the active spectrum contribution with sterile emission being kinematically disallowed. The magnitude of reduction throughout the entire spectrum is also dependent upon the size of the active-sterile mixing. 

\begin{figure}[t!]
	\centering
        \includegraphics[width=0.49\textwidth]{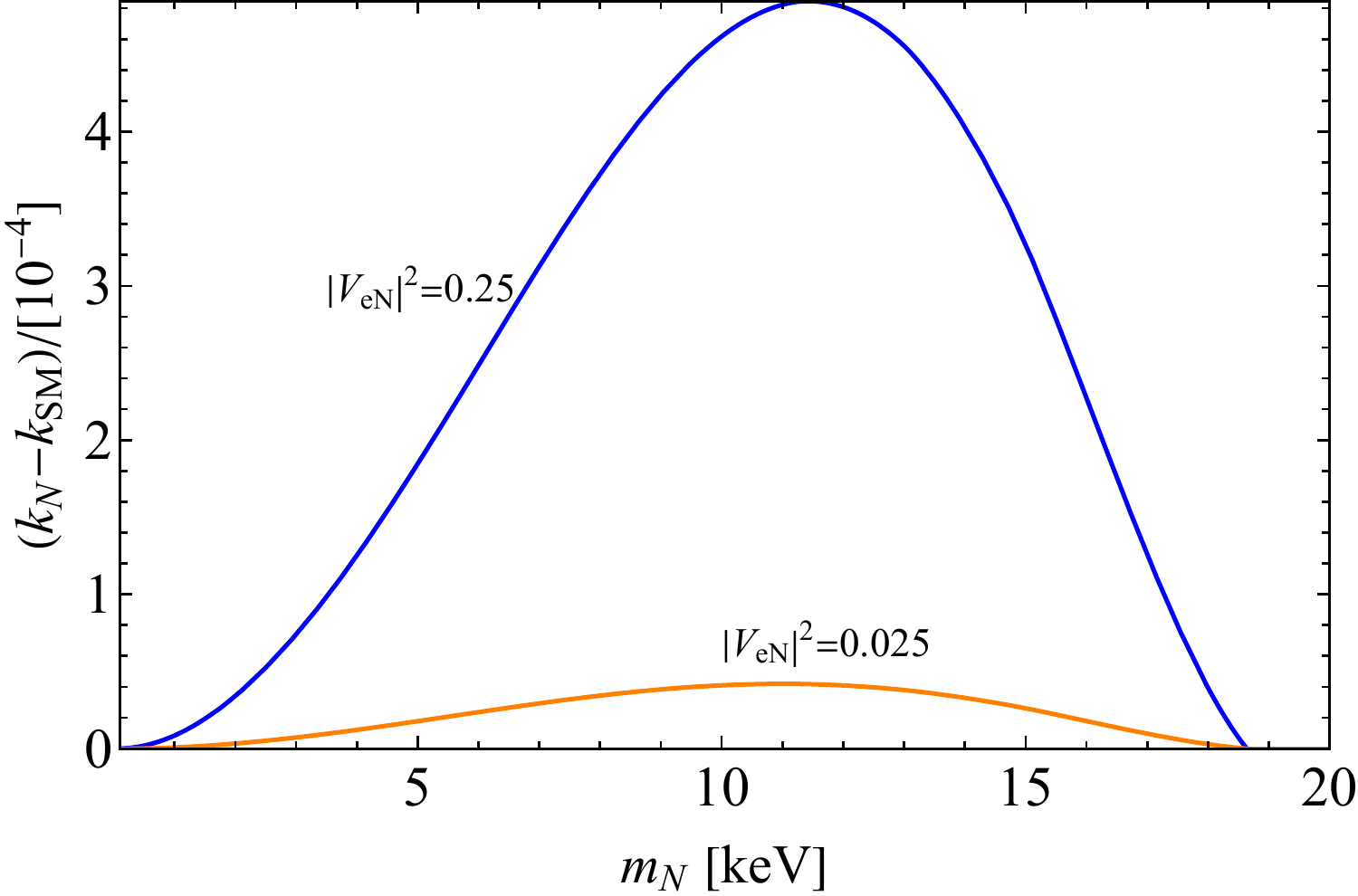}
	\includegraphics[width=0.49\textwidth]{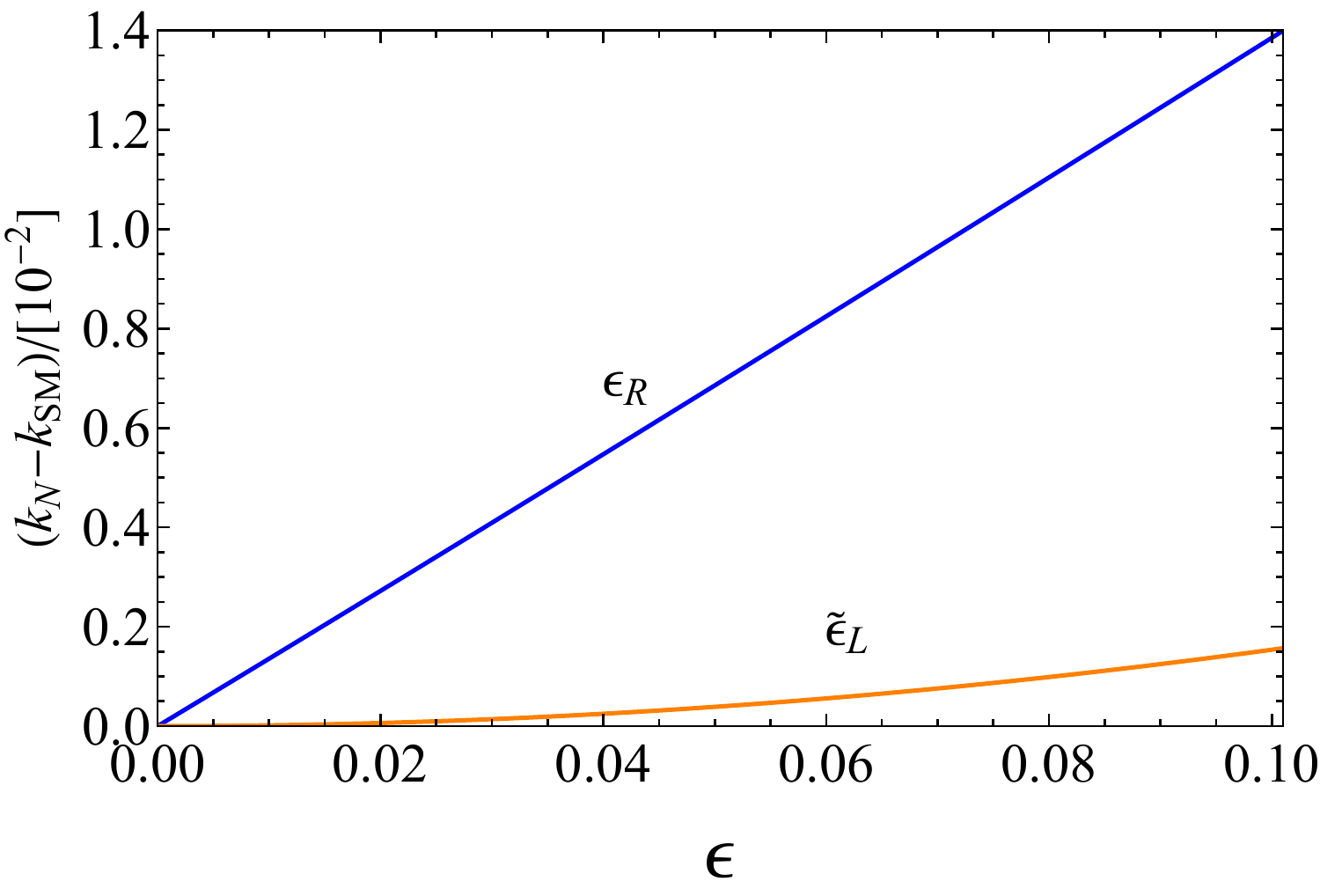}
	\caption{Deviation of the angular correlation factor $k$ from the SM value $k_\text{SM}$ as a function of the sterile neutrino mass for different values of the active-sterile mixing parameter $V_{eN}$ (left), and as a function of the exotic couplings $\epsilon_R$ and $\tilde{\epsilon}_L$ (right).}
	\label{fig:kvM}
\end{figure}

The impact on the angular distribution is depicted in Fig.~\ref{fig:kvM}~(left). In the case of a sterile neutrino only mixing with the active state and thus only inheriting a $V-A$ current, the effect is small and only arises from the impact of the larger sterile neutrino mass. For small sterile neutrino masses the angular correlation factor approaches that of the SM result. For large sterile neutrino masses the sterile neutrino becomes kinematically impossible to produce and thus its effect on the spectrum vanishes also. For intermediate values the heaviness of the sterile neutrino reduces the correlation between chirality and helicity (the cause of the anisotropy, as explained above) and reduces the magnitude of $k$, bringing the distribution closer to isotropy. For comparison, the impact of two exotic currents are also in Fig.~\ref{fig:kvM}~(right) where the deviation from the SM value increases as the parameters increase, although to differing extent. Note that the effect for the exotic currents is two orders of magnitude greater than that for active-sterile mixing.

However, we can also consider the presence of additional exotic currents that couple to the sterile neutrino directly from the Lagrangian of Eq.~\eqref{eq:ExLN}. These calculations are similar to those in the previous section, however as the final state is a different neutrino (sterile rather than active) there can be no interference between the SM result and currents involving a sterile neutrino,
\begin{align}
    a(E_e) &= a_\text{SM}(E_e) + |\epsilon^N_X|^2 a_X(E_e), \\
    b(E_e) &= b_\text{SM}(E_e) + |\epsilon^N_X|^2 b_X(E_e). 
\end{align}
The functions $a(E_e)$ and $b(E_e)$ are the same as for an active neutrino but using a different value for the neutrino mass and the relevant coupling $\epsilon^N_X$ is the one associated with the sterile neutrino.

\begin{figure}[t!]
    \centering
    \includegraphics[width=0.49\textwidth]{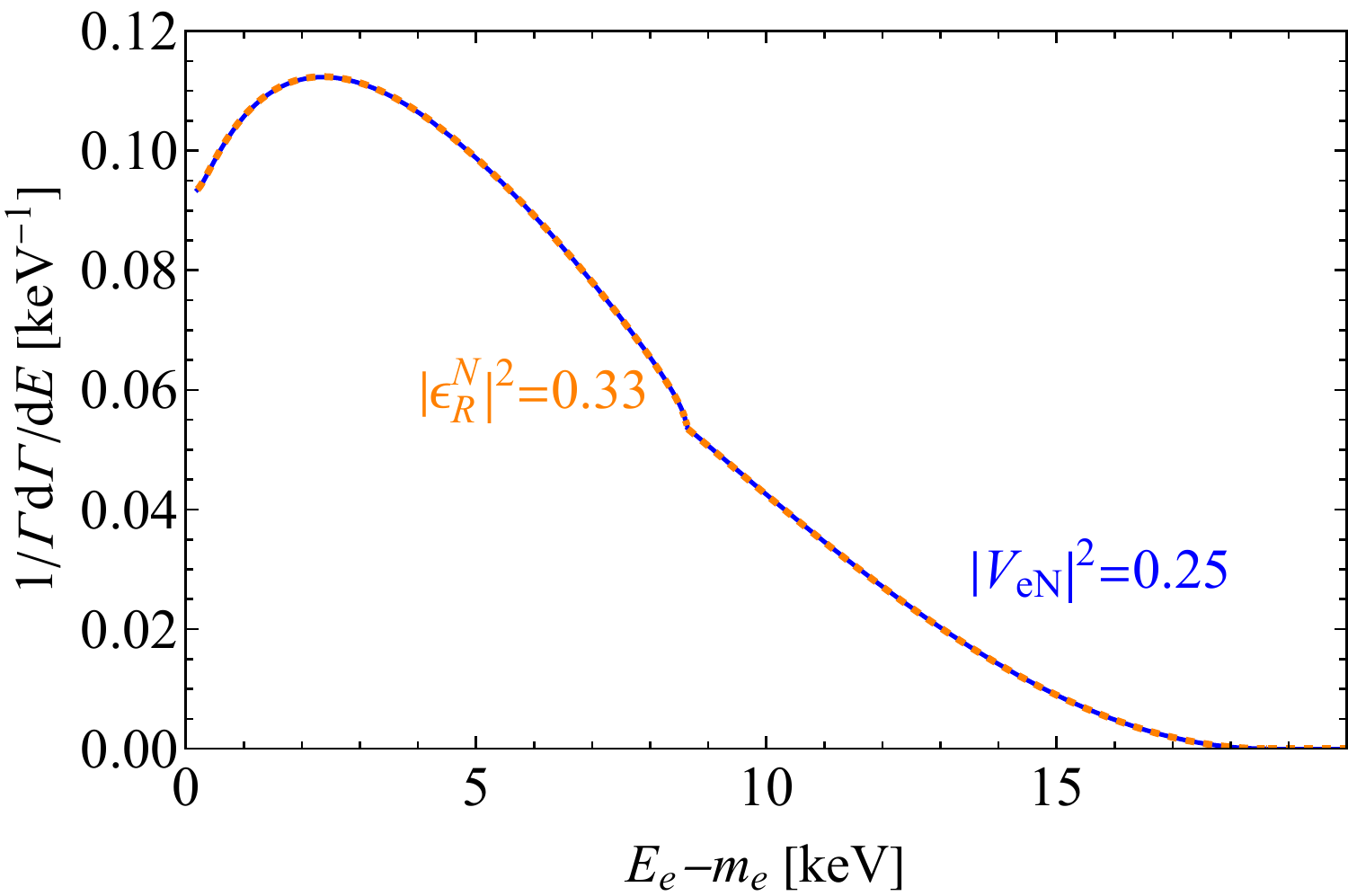} \includegraphics[width=0.475\textwidth]{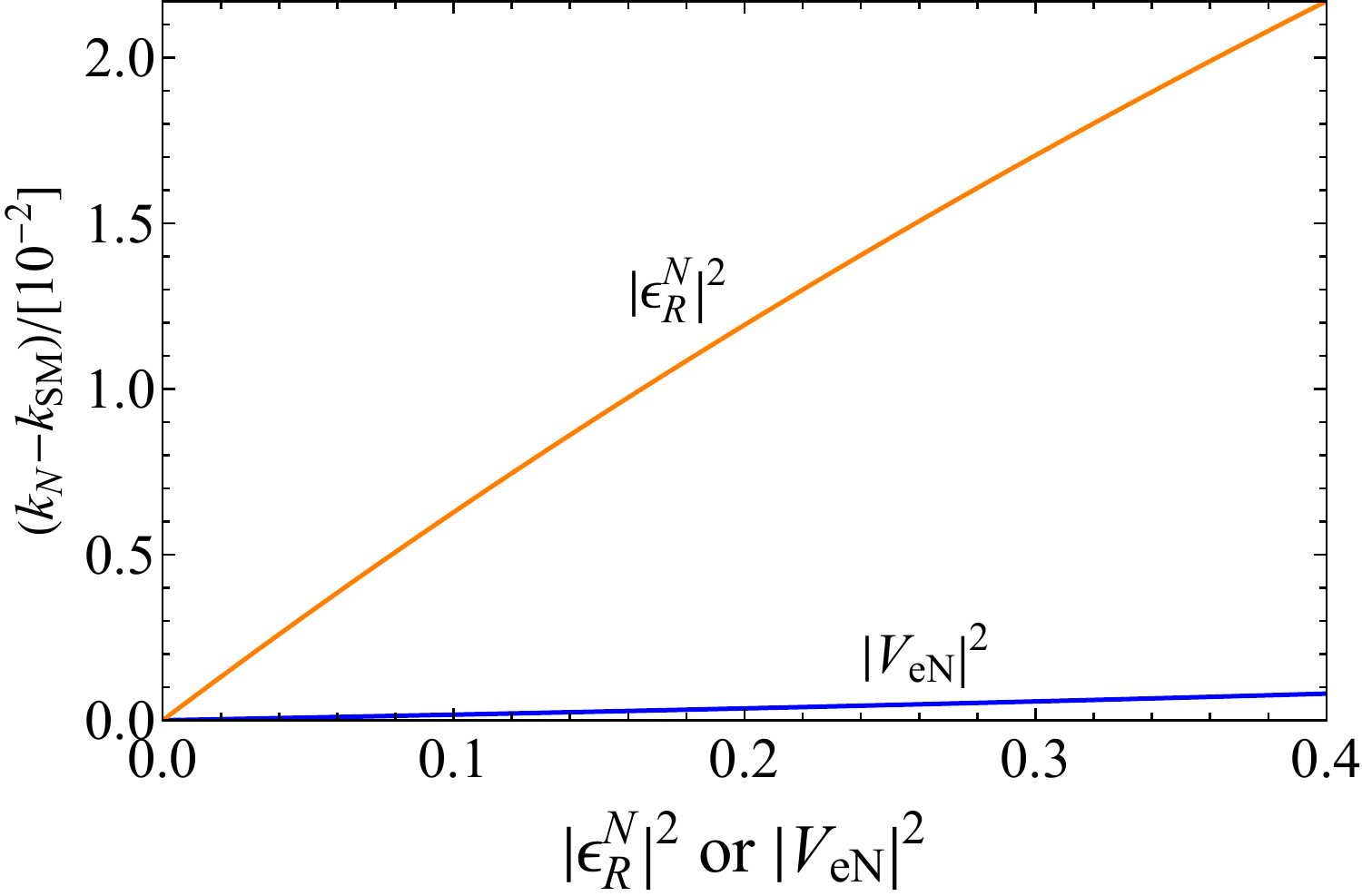}
    \caption{Energy and angular distributions showcasing the difference between an active-sterile mixing and an exotic right-handed leptonic current for producing a $10$~keV neutrino. Left: Normalised energy distributions with active-sterile mixing, $|V_{eN}|^2 = 0.25$ (blue, dashed), overlapping with an exotic sterile $|\epsilon^N_R|^2 = 0.33$ (orange) contribution. Right: Deviation of the angular correlation factor $k$ from the SM value as a function of the active-sterile mixing $|V_{eN}|$ or the right-handed leptonic current parameter $|\epsilon_R^N|$.}
    \label{fig:Ex vs Mix}
\end{figure}

In this way, we consider two mechanisms in which heavy neutrinos can be produced: indirectly, through the emission of an electron neutrino which has a heavy component because of active-sterile mixing, and directly, with an exotic current emitting a heavy sterile neutrino. Distinguishing between these scenarios could be difficult with some exotic currents matching the signatures of active-sterile mixing for particular values of $|\epsilon^N_R|^2$ and $|V_{eN}|^2$. As can be seen in Fig.~\ref{fig:Ex vs Mix}, in the energy spectrum both results are essentially identical, with a kink in each, due to the exotic current being approximately equivalent to the SM result. However, the angular correlation factor has a significantly different dependency on the parameters with the active-sterile mixing generating a much smaller variation. Simultaneous measurement of energy and angular spectra can aid in differentiating between these models as the degeneracies in parameter space occur at different points for the different scenarios.

\section{Tritium Decay Experiments} 
\label{sec:experimental}
The calculations in this paper have been carried out for both the energy spectrum and the angular distribution of the final state electron. Tritium $\beta$-decay experiments are mainly interested in measuring the absolute masses of the active light neutrinos and thus focus on the immediate endpoint of the energy spectrum.

There are three main experimental measurements of interest for the effects detailed here: precision measurements of the endpoint of the electron energy spectrum with the aim of measuring the active neutrino mass; measurements of the full spectrum, or a bulk segment of it, with sensitivity to any sterile neutrinos and exotic currents; and measurements of the electron angular distribution from a polarised source with sensitivity to exotic currents.

\subsection{KATRIN and TRISTAN} 
\label{sec:KATRIN}
The currently strongest upper bound on the effective neutrino mass is $m_\beta < 0.8$~eV at $90\%$ CL \cite{Aker2022} which exceeded the previous best limit of $m_\beta < 1.1$~eV \cite{aker2019improved, aker2020first}. The bound comes from the KATRIN experiment which uses a high-pass filter to measure the integrated electron kinetic energy spectrum focusing exclusively on the electrons within the final 300~eV near the endpoint \cite{Aker2022}. The KATRIN collaboration estimates their ultimate sensitivity to be $m_\beta \sim 0.2$~eV \cite{Huang:2019tdh}, hence there is great need for novel approaches if a guaranteed measurement of absolute neutrino mass is to be made, especially as the already large size of the central spectrometer (12~m diameter) makes scaling up to an even larger experiment - required for going beyond  0.2~eV - practically infeasible.

The TRISTAN project is a proposed extension of KATRIN with a novel detector and read-out system for a high-precision keV-scale sterile neutrino search, utilising KATRIN's gaseous molecular tritium source. The expected experimental signature of a sterile neutrino can be located anywhere along the electron kinetic energy spectrum, depending on the sterile mass. In contrast to this, the measurement of the effective active neutrino mass only distorts the end-point of the spectrum. Given that the objective of KATRIN is to constrain the latter, the apparatus has been designed in a way that accepts only the most highly energetic electrons, while filtering out the rest of the events. Therefore, KATRIN in its current state is not equipped to read out and handle the increased number of events, which would arise as a result of processing the entirety of the bulk, as opposed to just the end-point. TRISTAN has been proposed in order to use the KATRIN setup to probe sterile neutrinos by extending the acceptance range to cover the entirety of the spectrum, while also making use of upgraded equipment in order to allow for the resulting higher count rate to be processed. TRISTAN is expected to commence operation once KATRIN's data collection campaign has been completed, prospectively in 2025~\cite{Houdy:2020vhw}.

\subsection{Project 8}
The Project 8 collaboration aims to directly measure the mass of electron neutrinos making use of Cyclotron Radiation Emission Spectroscopy (CRES) technology \cite{Pettus:2017sxd}. When an emitted $\beta$-electron passes through a magnetic field, its circular acceleration causes it to emit cyclotron radiation. The frequency $f$ of this radiation depends on the energy of the electron,
\begin{align}
	f = \frac{1}{2\pi}\frac{eB}{E_e},
\end{align}
where $e$ is the electron charge, $B$ the magnetic field strength and $E_e$ the total electron energy. Precise measurements of the frequency (using technology already developed) thus allows for a high energy resolution, provided that the magnetic field is sufficiently well known \cite{Pettus:2017sxd}. Measurement of the energy spectrum can thus take place with attention being paid to either the endpoint or the full spectrum. The Project 8 collaboration proposes to use this technology with the target of placing an upper bound of $m_\beta \lesssim 40$~meV \cite{Pettus:2017sxd} but with a target energy resolution of roughly $100$~eV when looking across the whole spectrum~\cite{Drewes2017}. Project 8 have estimated that the limit in sensitivity from a molecular source of tritium has an irreducible value of $\approx 0.3$~eV \cite{monreal2012project} as a result of the Final State Distribution (FSD) problem, which arises due to final state excitation of molecular tritium. This is not accurate enough to guarantee a measurement of the neutrino mass given the `worst'-case minimum value of $m_\beta \sim 9$~meV for normal ordering and $m_\beta\sim 40$~meV for inverted ordering \cite{doe2013project}. A solution to this is to use atomic instead of molecular tritium, resulting in the achievable sensitivity increasing by a factor of 2 (or even more, given a large enough effective volume), which is what allows Project 8 to project their sensitivities probing as low as 40~meV.

\subsection{CRESDA}
Although the use of atomic tritium as proposed by the Project~8 collaboration is expected to greatly improve the sensitivity, it is not sufficient if the neutrino mass realised in nature is smaller than $\approx 40$~meV. Therefore, other probes are required if a guaranteed measurement is to be made. In an attempt to increase the energy resolution of measurements exploiting CRES technology, and hence to push the existing and proposed bounds even further, the \emph{Quantum Technologies for Neutrino Mass} (QTNM) consortium, consisting of several UK institutes has been formed with a proposal to make guaranteed direct neutrino mass measurements with the help of quantum technologies~\cite{QTNM:2020}. First, the aim is to develop the experimental tools and technologies using hydrogen and deuterium atoms before moving on to tritium - this is the CRES demonstration apparatus (CRESDA). The main challenges of a CRES measurement that the current phase of the QTNM proposal aims to address are the production and confinement of tritium atoms, mapping magnetic fields in the CRES detection region with high precision, and the realisation of high sensitivity microwave electronics devices for detection and characterisation of the cyclotron radiation. Over the 3 year timeline, the goal of CRESDA is four-fold: to produce and confine large quantities of hydrogen and deuterium atoms at densities exceeding $10^{12}~\text{cm}^{-3}$; use the hydrogen/deuterium atoms as quantum sensors to implement Rydberg atom magnetometry techniques to map the magnetic field with a precision on the order of $\pm 100$~nT; to develop microwave detection and readout electronics components to operate at frequencies on the order of 10~GHz, and allow a spectral resolution at the ppm level; and to design a software framework for simulations, signal processing and sensitivity projections.

\subsection{Future Sensitivity Goal}
\begin{figure}[t!]
	\centering
	\includegraphics[width=0.75\textwidth]{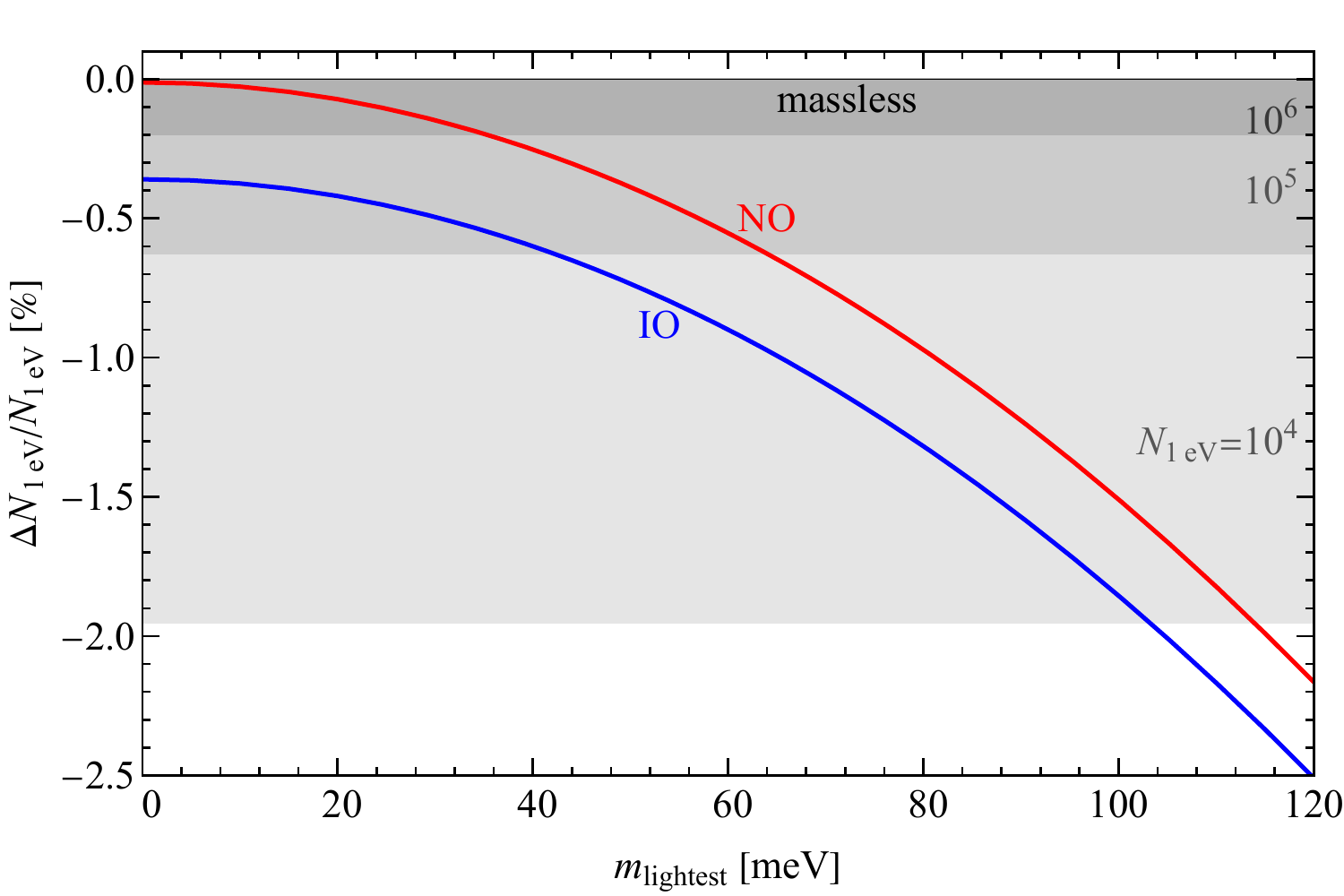}
	\caption{Percentage variation in the number of events in the final 1~eV of the tritium energy distribution as a function of the lightest active neutrino mass. The red line denotes the normal ordering case and the blue line the inverted ordering. The gray bands denote the 95$\%$ CL for a detectable change in the rate as a function of the total number of events in the final eV.}
	\label{fig:ChangeInN}
\end{figure}
In this work, we aim to determine the sensitivity of future tritium decay experiments to exotic currents and to keV-scale sterile neutrinos. This clearly depends on the details of the experiments but in order to estimate the sensitivity we assume an exposure such that the experiment can in principle distinguish between the normal order and inverse order of active neutrinos when measuring the neutrino mass near the end-point. In a $\beta$-decay experiment of the type considered in this paper, the $\beta$-decay tritium source produces a number of events $N_\text{total}$ over the observation time. After experimental acceptance and efficiency corrections, collectively taken into account through a factor $\eta \leq 1$, a number of events is collected, $N_\text{events} = \eta N_\text{tot}$ (throughout this work we take $\eta=1$). In the analysis of, e.g., the light neutrino mass, events are specifically considered in a window,
\begin{align}
    N_\text{win} = \frac{N_\text{events}}{\Gamma}
    \int_{E_1}^{E_2} \frac{d\Gamma}{dE_e}dE_e,
    \label{Nwin}
\end{align}
where $E_1$ and $E_2$ are the upper and lower limits of the energy window under consideration, respectively. The normalization is with respect to the total tritium decay rate.

In the determination of the light neutrino mass in future experiments, a focus will be on the final 1~eV below the endpoint. Fig.~\ref{fig:ChangeInN} shows the relative change in the number of events in the last 1~eV of the spectrum as a function of the lightest active neutrino mass, $m_\text{lightest}$, for NO and IO cases. In principle, $m_\text{lightest} = 0$ is possible, therefore one needs to consider the worst case scenario when calculating the number of $\beta$-decay events needed to resolve the two hierarchical cases. One can estimate the sensitivity via the fluctuation of the number $N_\text{1eV}$ of events in the final eV, $\sigma_N = 1/\sqrt{N_\text{1eV}}$. It can be seen that the percentage deviation for a massless lightest neutrino in the IO case is $-0.37\%$. For the 95\%~CL bound in a one-tailed test requires a minimum number of events of $N_\text{1eV} \approx 2.8\times 10^5$. Using Eq.~\eqref{Nwin}, this corresponds to a total number of events of $N_\text{tot} = 1.3 \times 10^{18}$. In order to derive our sensitivities, we will assume a default number of total events of $N_\text{tot} = 10^{18}$.

The TRISTAN experiment is aiming to reach a sensitivity of $\abs{V_{eN}}^2 \lesssim 10^{-6}$ on the active-sterile neutrino mixing in the future, assuming $10^8$ counts per second (cps) over a 3~year data taking period. This corresponds to a total of $10^{16}$ events. It is estimated that the experiment could reach the statistical limit with 3~years of data taking and by utilising the maximal source strength of KATRIN, $10^{10}$~cps, corresponding to $10^{18}$ events. This matches with our default value for $N_\text{tot}$.

\subsection{Impact of Angular Correlations}
In order to perform measurements of the angular distribution the tritium nucleus must be polarised and the emission direction (or at least its hemisphere relative to the polarisation) must be measurable. There has already been much interest in polarising hydrogen nuclei due to their potential use in nuclear fusion reactions \cite{Hupin:2019}. An Atomic Beam Source (ABS) is used for ANKE at COSY-J\"ulich to produce a $\sim20$ K hydrogen beam with a $95\%$ disassociation (i.e. proportion of atomic hydrogen) and $\sim90\%$ polarisation \cite{Mikirtychyants:2012nh}. The polarised hydrogen can then be put in a storage cell with measurements finding that on the order of hours there was no decay of the polarisation \cite{Ciullo:2011}.

Such a beam could not directly be the source for a Project 8 like experiment which requires a fraction $T_2/T\leq10^{-6}$ and temperature $130-170$ mK in order to perform its precision endpoint measurements \cite{Project8:2017nal}. The Project 8 experiment expects to use a strong magnetic trap filled with ${}^4$He to both prevent molecular recombination and cool the tritium. At the low pressures of the experiment, virtually all of the recombination will occur at the walls; using a strong magnetic field will trap the T, using its magnetic moment, whilst allowing any T$_2$, which lacks a magnetic moment, to escape. The gas of ${}^4$He, also lacking a magnetic moment, could be used to maintain contact with the walls of the vessel and cool the tritium \cite{Project8:2017nal, Clark:2014}.

Measuring both the energy and emission direction of the electron is expected to be challenging. However, the measurements suggested herein require only knowledge of the hemisphere of emission relative to the tritium spin. Thus if the tritium is moving sufficiently rapidly at the time of its decay, with its momentum parallel to its spin, the emission hemisphere of the electron could be inferred from its kinetic energy provided the boost from being emitted parallel to the tritium motion is larger than the energy provided by the decay. A cut could be performed in the data to remove any emissions ambiguously on the boundary between the two hemispheres. Alternatively, the detection antenna could be placed in the direction of the spin and whether the emitted electron was headed towards or away from the antenna determined. Crucially, the direction of nuclear polarisation and the external magnetic field are independent and thus so is the electron `pitch angle' and `polarisation angle'. Hence there is freedom in choosing the polarisation direction relative to the magnetic field to maximise the sensitivity to the electron emission hemisphere.

Overall, designing an experiment that can perform all three measurements (end-point, full spectrum and emission angle) is expected to be very challenging due to the high precision required for setting new limits on the light neutrino mass. However, performing full and end-point spectrum measurements is within the target design of the Project 8 experiment and existing polarisation technologies could be implemented within a similar setup to measure the angular and energy spectra. Tritium is not well suited to probe the angular spectrum, as the $Q$ value is low leading to a small kinetic energy compared to the electron mass, leading to a washout of deviations. We nevertheless discuss the prospect of probing it as tritium searches such as Project~8 are expected to become crucial for neutrino physics.

\section{Future Sensitivity to New Physics} \label{sec:Sensitivity}

\subsection{Statistical Analysis}
So far the presence of sterile neutrinos and exotic currents has gone unobserved. This lack of observation has placed bounds on the parameters that quantify these effects, $V_{eN}$, $m_N$ and the $\epsilon_X$. Sensitivity to New Physics can be calculated using a minimal-$\chi^2$ test with binning for either the energy spectrum or the angular distribution. For the angular distribution it is appropriate to use two bins representing the `aligned' and `anti-aligned' hemispheres. For the energy spectrum the number of bins is merely limited by the energy resolution of the experiment provided that the bin width is larger than said resolution \cite{Huang:2019tdh}.

In order to reflect the uncertainty in the overall normalisation of the energy spectrum, a nuisance parameter, $A$, on the overall expected rate is introduced. Thus our test statistic is calculated using
\begin{align}
	t = \text{min}_A
	\left[\sum_{i=1}^{N_\text{bins}}\frac{(N_\text{BSM}^{(i)}-(1+A)N_\text{SM}^{(i)})^2}{N_\text{SM}^{(i)}}
	+ \left(\frac{A}{\sigma_A}\right)^2\right],
\label{eq:teststat}
\end{align}
where the subscript `BSM' denotes the contribution associated with New Physics, with the corresponding parameter of interest, such as $\abs{V_{eN}}^2$, $m_N$ or any of the exotic currents considered, $\epsilon_X$. The uncertainty on the nuisance parameter $\sigma_A$ is set to $\sigma_A = 2$ (conservatively large but with little impact). In Eq.~\eqref{eq:teststat}, $t$ denotes the test-statistic and is the value we use to measure the deviation from the SM. Theoretically, the expected number of events needs to be the median of the expected events over a large number of runs, as every run will produce a slightly different number of observed events, leading to statistical fluctuations. This can be simulated by running a series of Monte Carlo tests, which is usually a computationally rather expensive task. Instead, one can use the Asimov data-set, which allows for the median values over many runs to be replaced by their expectation values \cite{gcowan2011}.

Throughout this paper the number of bins for the energy spectrum is set to twenty (apart from Fig.~\ref{fig:constraints}, where the bin width is 300~eV, comparable with the choice by TRISTAN). In the asymptotic limit, which is assumed in this paper due to the very large number of events, the number of events in each bin can be replaced by the expectation values, as outlined above. Thus, the content of each bin is a fixed number, rather than a randomly fluctuating distribution, which removes the degrees of freedom associated with the bins themselves. As a result, the log-likelihood tends to a $\chi^2$ distribution with one degree of freedom per strength parameter (i.e., $|V_{eN}|^2$ or one of the $\epsilon$). At the $95\%$ CL, a $\chi^2$ in excess of $3.841$ for one degree of freedom and $5.99$ for two degrees of freedom is significant. Limits on the corresponding parameters within the `observed' model can thus be placed at the 95\%~CL if deviations from the SM are not observed. The $\chi^2$ has been calculated with an overall number of events of $N_\text{tot} = 10^{18}$ across all bins which corresponds to a total of $\approx 2.8\times10^5$~events in the final eV below the endpoint. 

\subsection{Exotic Currents}

In future $\beta$-decay experiments, measurements of the energy or angular spectrum will allow one to either obverse or place limits upon exotic currents. Here we summarise the upper bounds that can be placed on the $\epsilon$ parameters in the absence of a deviation from the SM expectation. These are for an ideal experiment, without systematic uncertainties, with a total expected number of $10^{18}$ events. 

\begin{figure}[t!]
	\centering
	\includegraphics[width=0.9\textwidth]{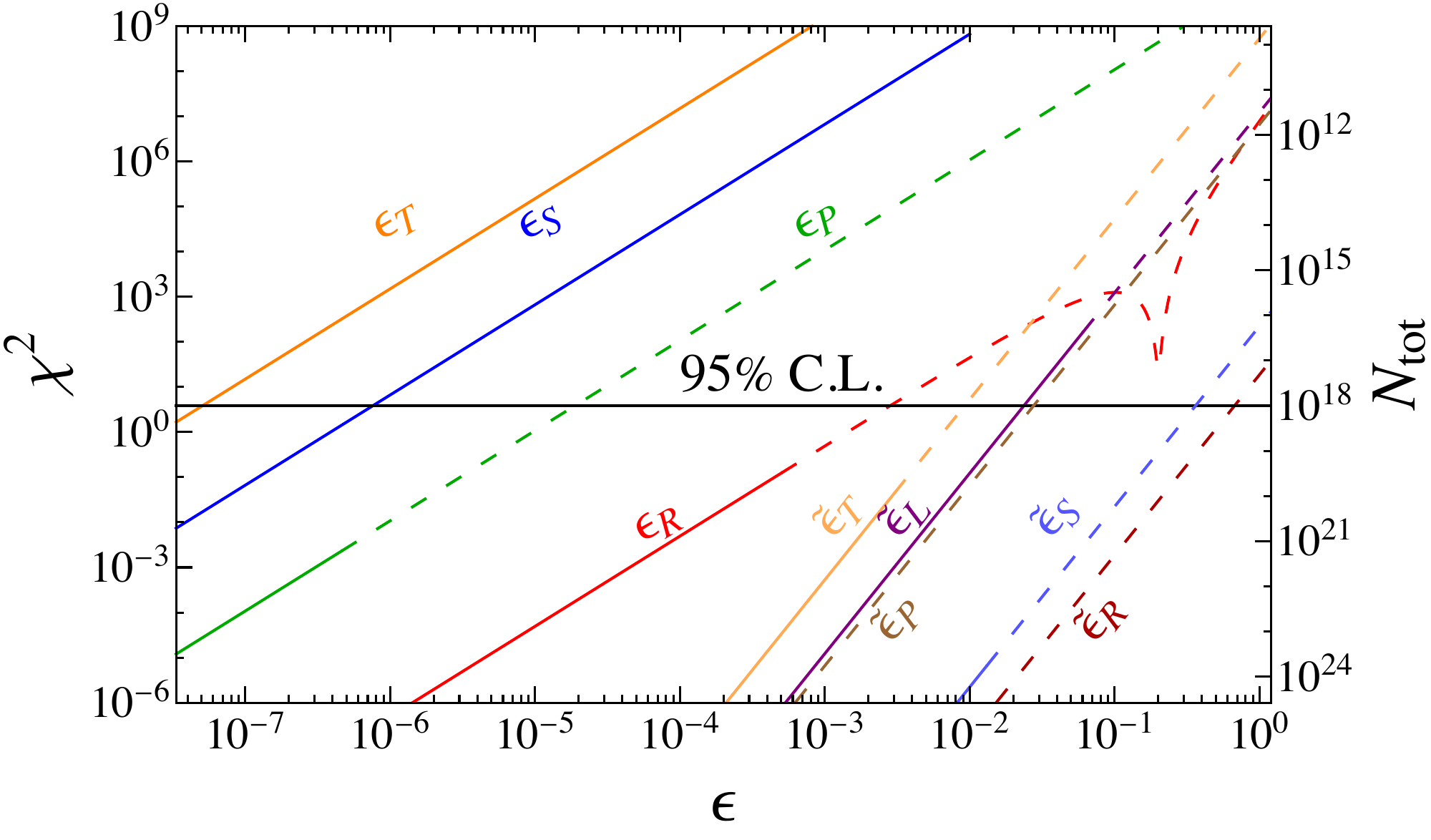}
	\caption{$\chi^2$ deviation for exotic current contributions as a function of the relevant coupling constant, using the tritium energy distribution. The curves are plotted in solid below the coupling constant's upper bound (cf. Table~\ref{tab:upper-bounds}) and dashed above. The horizontal black line is the 95$\%$~CL sensitivity for $N_\text{tot} = 10^{18}$ events, while the $N_\text{tot}$ scale on the right-hand axis shows the number of events required to reach a 95\%~CL sensitivity. The labels denote the relevant coupling constant contributing at a time.}
	\label{fig:ExP}
\end{figure}
Fig.~\ref{fig:ExP} gives the $\chi^2$ deviation when individual exotic currents are included as a function of their corresponding parameter. As can be seen, the sensitivity to scalar and tensor currents is the greatest and much more significant than that to pseudoscalar currents. This is due to the functional form of the additional terms with the pseudoscalar current being functionally suppressed. For the right-handed leptonic currents, the linear term is proportional to the neutrino mass, and thus negligible, meaning that the quadratic term is the main contribution. This gives these currents a different gradient to the others as the $\chi^2$ for these is dependent upon $\epsilon^4$ unlike the others which go like $\epsilon^2$. The energy sensitivity to the vector currents is much less as the additional terms are nearly identical to the SM expression (see Appendix~\ref{App:full-expressions}) and thus sensitivity to them is obscured by the uncertainty in the overall normalisation (i.e., the minimisation over $A$). The dip in the right-handed hadronic current comes from the point at which the linear and quadratic terms are roughly comparable in magnitude leading to partial cancellation. For values of $\epsilon$ above this, the quadratic term dominates and the right-handed hadronic term tends towards being the same as the right-handed leptonic term.

For the angular distribution, the dependencies are largely the same but with some currents experiencing much greater sensitivity. This of course assumes an experiment that uses perfectly polarised tritium and that can probe the angular distribution with no loss of efficiency. For the scalar and tensor currents the primary variation comes from the energy-dependent change in the energy spectrum (and thus the overall rate) which is not obscured by the overall normalisation. This is the primary effect measured in both the energy and angular distributions thus giving them the same sensitivity. However, for the pseudoscalar current the energy dependent change is highly suppressed so for the angular distribution the additional angular term is most significant. For all of the right-handed terms the normalisation uncertainty removes most of the energy sensitivity but cannot simultaneously remove the angular sensitivities which have a different dependence thus they are much more significant. On the right-hand scales in the two figures we note the required number of events to give a $95\%$~CL sensitivity. This follows inversely proportional to the $\chi^2$.
\begin{figure}[t!]
	\centering
	\includegraphics[width=0.9\textwidth]{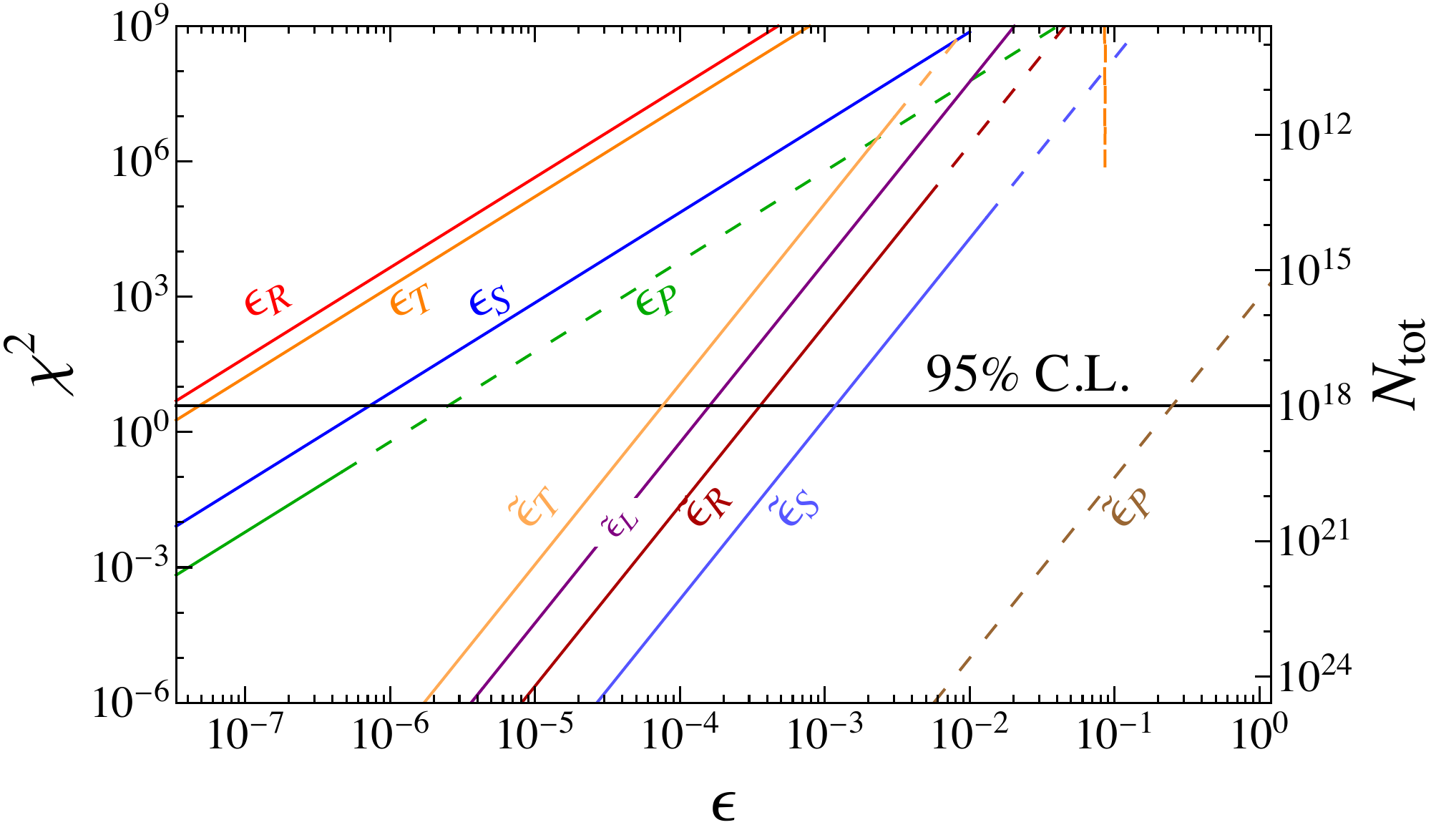}
	\caption{As Fig.~\ref{fig:ExP} but using the angular distribution.}
	\label{fig:AngExP}
\end{figure}
\begin{figure}[t!]
	\centering
	\includegraphics[width=0.49\textwidth]{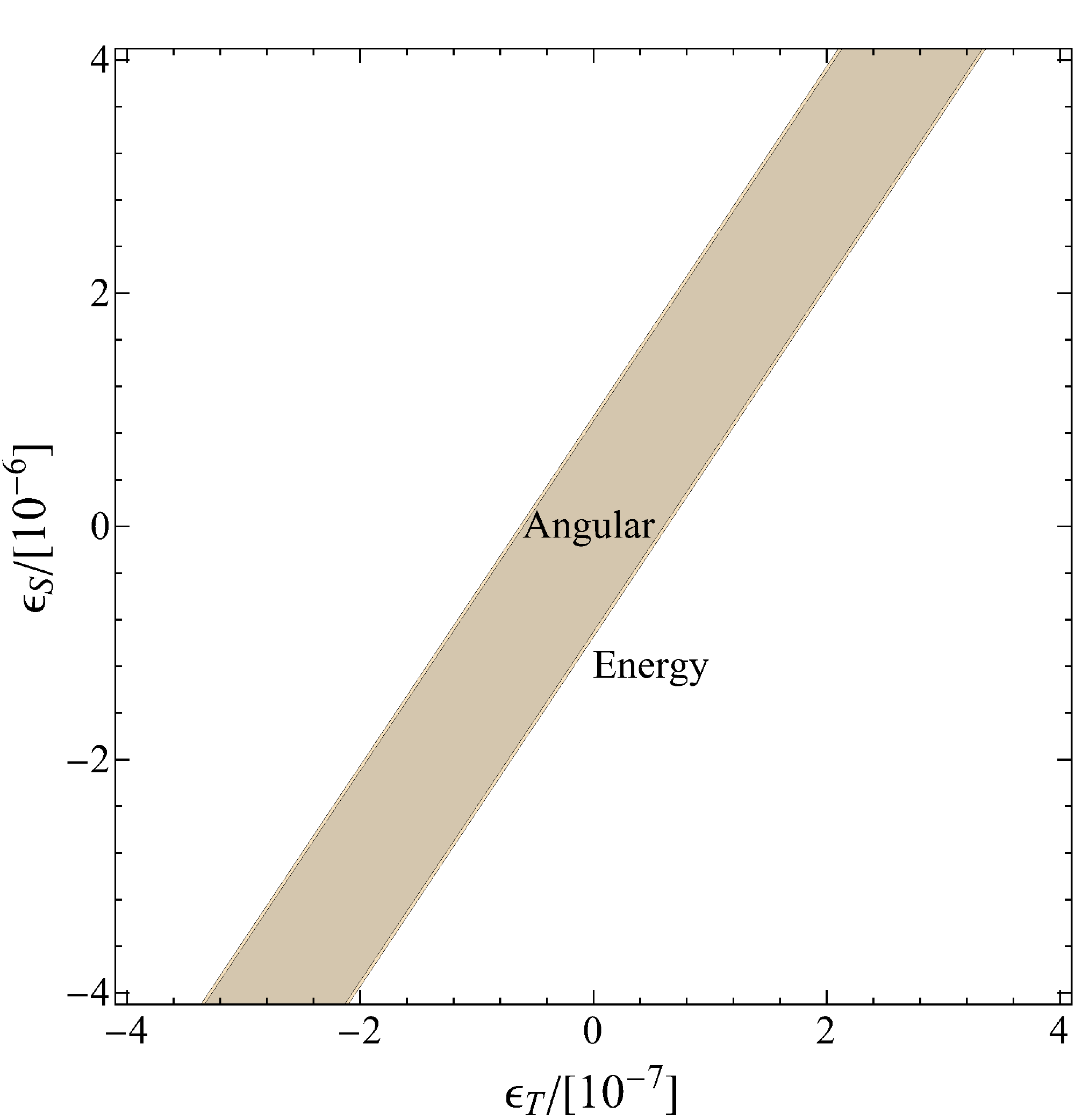}
	\includegraphics[width=0.49\textwidth]{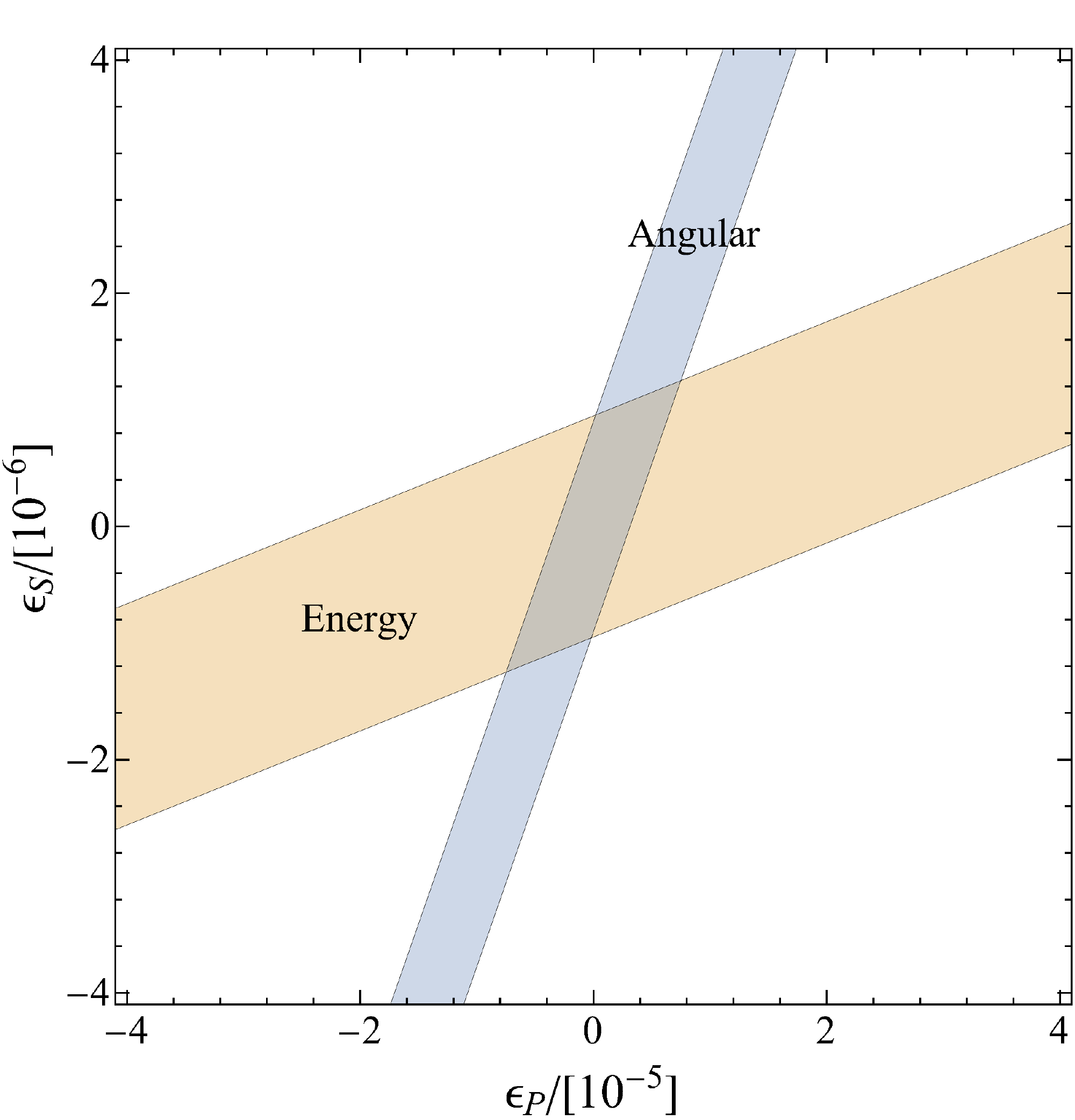}\\
	\includegraphics[width=0.49\textwidth]{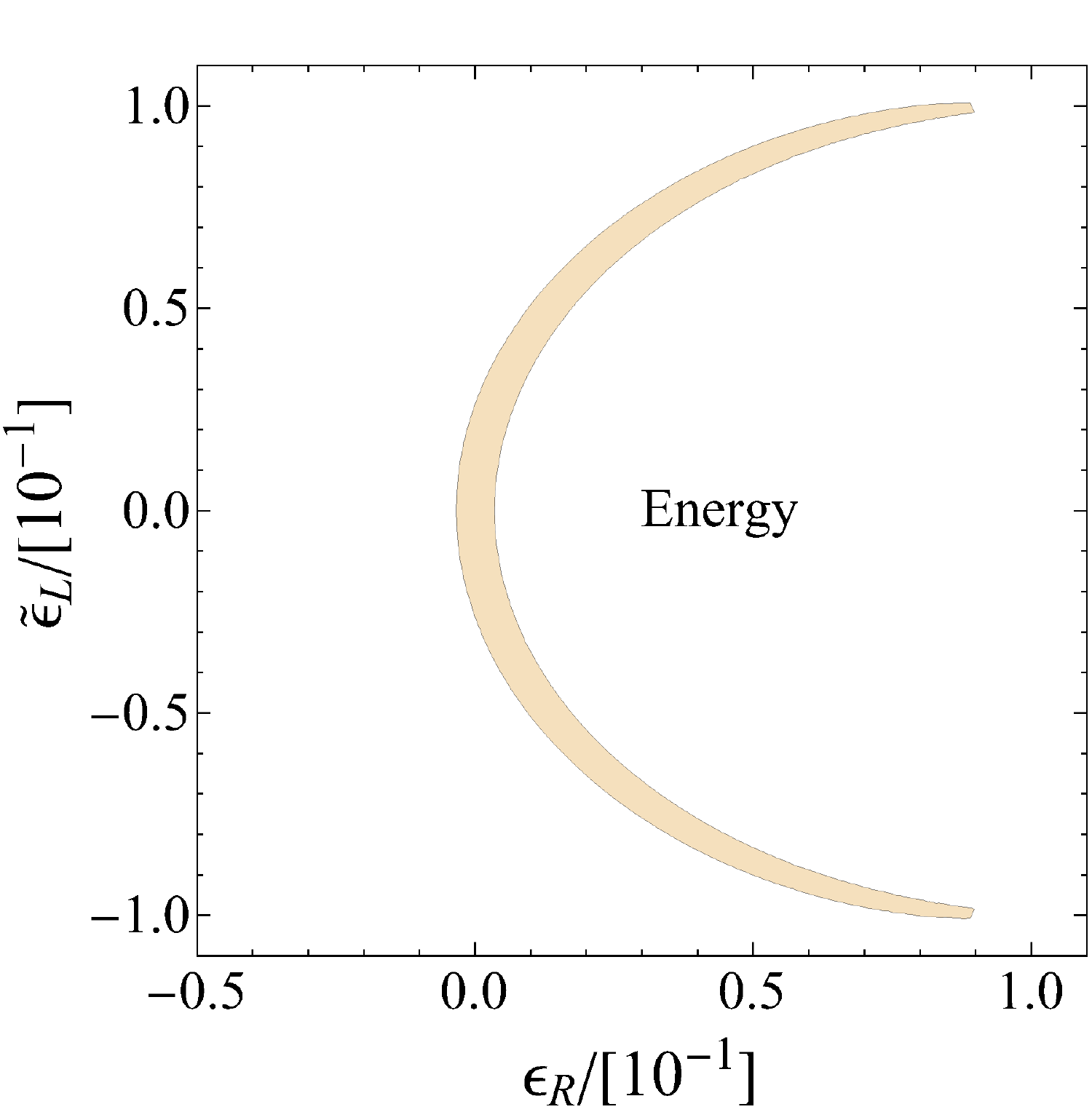}
	\includegraphics[width=0.49\textwidth]{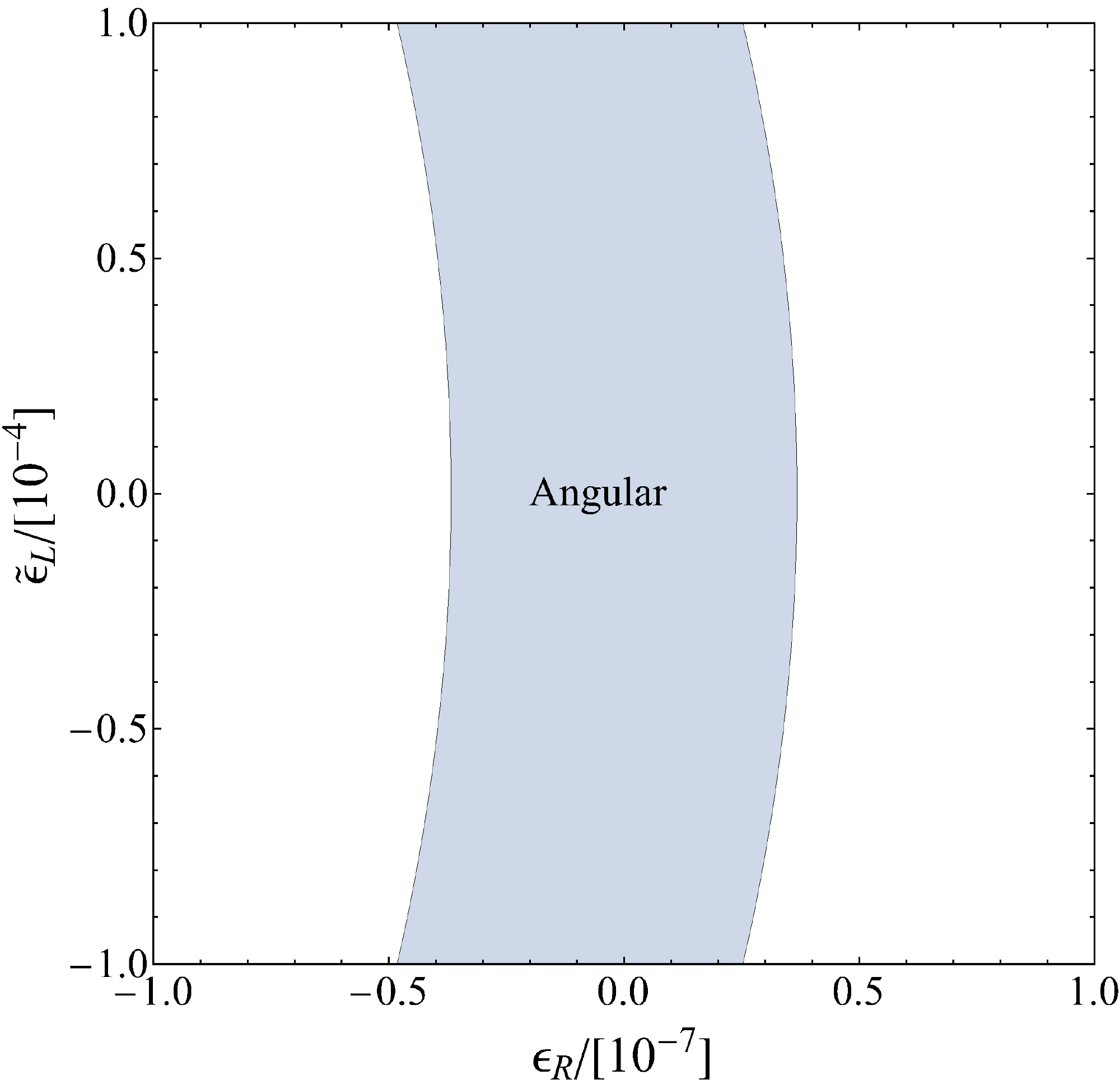}
	\caption{95$\%$~CL regions for BSM contributions driven by two exotic currents for $N_\text{tot} = 10^{18}$ events. The region based on the energy (angular) distribution is in orange (blue). The combinations of currents are: scalar and tensor (top left), scalar and pseudoscalar (top right), right-handed hadronic and right-handed leptonic (bottom, left for energy and right for angular).}
    \label{fig:ExComp}
\end{figure}
In actual New Physics scenarios it is unlikely that exotic currents are triggered individually as they all potentially contribute to the energy and angular distributions. A result that is compatible with the SM in these measurements is, for example, also compatible with the simultaneous presence of scalar and tensor currents. This can be seen in Fig.~\ref{fig:ExComp} where 95$\%$~CL exclusion contours are shown for pairs of parameters. In the scalar, pseudoscalar and tensor cases the additional linear terms dominate so the regions appear to be straight lines of cancellation. For the scalar versus tensor plot the difference between energy and angular distribution is negligible. However, the additional sensitivity of the angular pseudoscalar distribution leads to different gradient lines in the scalar vs pseudoscalar plot. For the right-handed leptonic current the dominant contribution is quadratic unlike the linear dominance for the right-handed hadronic. This leads to parabolic regions for small values but which become more complicated for larger values where the quadratic part of the right-handed hadronic current becomes relevant. 

\subsection{Sterile Neutrinos}
We can apply the same treatment as above to look at the sensitivity to production of a sterile neutrino. This neutrino could either mix with the active states or be produced directly through exotic currents. Thus we can place limits on the mixing, $V_{eN}$ or the parameters $\epsilon^N$ which will in general be dependent upon the mass of the (mostly) sterile state. 

\begin{figure}[t!]
    \centering
    \includegraphics[width=0.8\textwidth]{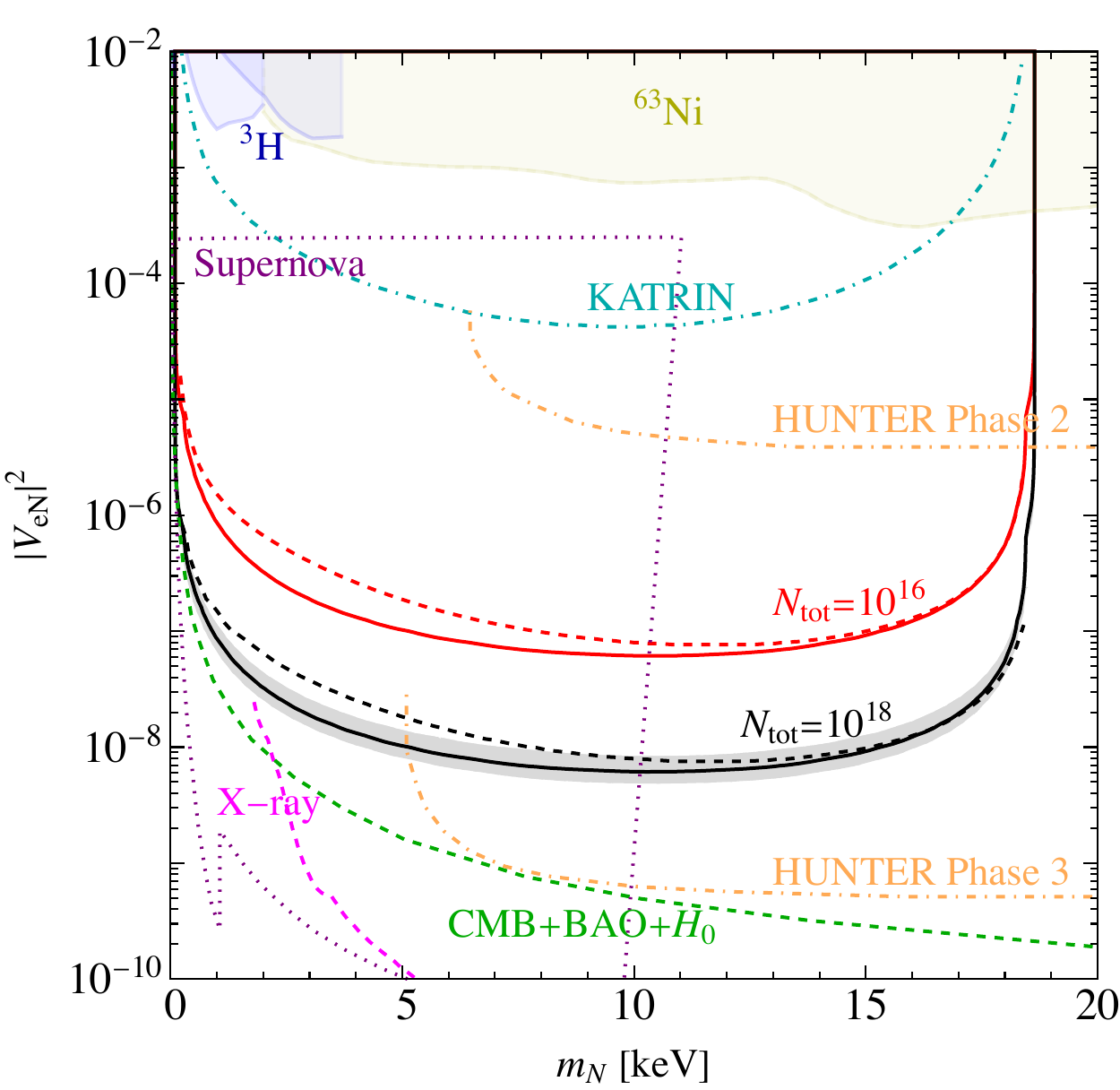}
    \caption{Projected sensitivities to the active-sterile mixing $|V_{eN}|^2$ as a function of sterile neutrino mass $m_N$ at 95\%~CL, for a total number of events $N_\text{tot} = 10^{16}$ (red solid) and $N_\text{tot} = 10^{18}$ (black solid), alongside corresponding expected sensitivities of TRISTAN (dashed red and dashed black). The gray shaded band corresponds to the $1\sigma$ variation of the $95\%$~CL for a large number of simulations. The shaded regions are excluded by $^{3}$H (blue) and $^{63}$Ni (yellow) searches together with future experimental constraints expected from KATRIN (cyan dot-dashed) and HUNTER (orange dot-dashed). The dotted lines show current astrophysical constraints from: X-ray data (pink) \cite{ng2019new}, CMB+BAO+$H_0$ observations (green) \cite{vincent2015revisiting} and supernova data (purple) \cite{shi1993type}.}
    \label{fig:constraints}
\end{figure}
Considering first, the case of a one-active + one-sterile neutrino gives a sensitivity to mass and mixing as shown in Fig.~\ref{fig:constraints}. The impact of the sterile neutrino is most felt through the `kink' in the energy spectrum which occurs at an energy equal to the sterile mass below the endpoint. As can be seen, the greatest sensitivity to mixing is at sterile masses around half of the maximum kinetic energy of the electron. For low sterile masses the sensitivity is reduced because the sterile spectrum looks nearly identical to the active spectrum. For high sterile masses the sterile spectrum increasingly vanishes as the neutrino cannot be kinematically generated, this merely leaves a reduced active spectrum but with the overall rate reduction hidden by the uncertainty in the normalisation of the SM spectrum (i.e. the minimisation over $A$ within the $\chi^2$, see Eq.~\eqref{eq:teststat}).

In Fig.~\ref{fig:constraints}, the tritium sensitivity is compared to existing bounds (shaded regions) and projected sensitivities from other experiments as well as (model-dependent) astrophysical bounds. As can be seen, the projected limits are clearly stronger than any existing constraints from direct $\beta$-decay measurements or even the projected sensitivity from KATRIN \cite{mertens2019novel} over the considered sterile mass range. The solid black (red) curve corresponds to the projected sensitivity based on our calculations, for a total number of events $N_\text{tot} = 10^{18}$ ($N_\text{tot} = 10^{16}$). For comparison, the corresponding projected sensitivity of TRISTAN \cite{mertens2019novel} is given by the dashed black and red curves.

Astrophysical and cosmological considerations can be used to constrain the presence of sterile neutrinos in the universe. Sterile neutrinos are viable dark matter candidates, which could have been produced in the early universe by mixing with the active neutrinos. If they exist, the heavy states can decay into an active neutrino alongside a mono-energetic photon, producing a distinct line in the X-ray domain. Data from NuSTAR is thus able to constrain the mass and mixing of a new heavy sterile state~\cite{ng2019new}. A heavy neutrino state which decouples between Big Bang Nucleosynthesis (BBN) and recombination can make the Universe appear to be younger and hence result in a larger Hubble parameter. This effect can be balanced out by the decay of the heavy states into neutrinos. The CMB and the sound horizon from Baryon Acoustic Oscillations (BAO) which are remnants from the recombination era can be used to put bounds on the mass and mixing of such a heavy state~\cite{vincent2015revisiting}. Active-sterile neutrino mixing can have profound effects in supernova explosions. A rather large mixing may disable a supernova from exploding, while in the case of a proto-neutron star, the oscillation can lead to a cooling effect and thus decrease the cooling time, which has been observed to be of the order of several seconds. These potential effects constrain the allowed active-sterile mixing~\cite{shi1993type}.

Another future search, HUNTER, will utilise the decay of cesium following electron capture to measure the masses of active and sterile neutrinos. In Phase 1, which is currently undergoing construction, the source will comprise of $10^8$ atoms, resulting in $2.1\times10^5$ events per 360 days of active running. However, this will only cover a mass range of approximately 50~keV to several 100~keV, and hence lies outside of the relevant parameter space for this work. In Phases 2 and 3, HUNTER will operate with $4\times 10^9$ and $3\times 10^{11}$ source atoms, resulting in $4.3\times 10^7$ and $8.7\times 10^{10}$ events per 360 days of active running, respectively. In terms of its sensitivity to the active-sterile mixing angle, HUNTER is projected to surpass $10^{-5}$ in Phase 2, and is aiming to reach $10^{-9}$ in Phase 3 \cite{martoff2021hunter}. HUNTER Phase 2 is expected to be significantly less sensitive than TRISTAN or a new future search based on our calculations, while Phase 3 (as well as Phase 2) will not cover the mass range below approximately 5 keV. Therefore, despite the promising projections for HUNTER Phase 3, a separate search is required in the 0-5 keV range. The constraints from X-ray, cosmological and supernovae data are in places more sensitive but are significantly more model-dependent.

\begin{figure}
	\centering
	\includegraphics[width=0.8\textwidth]{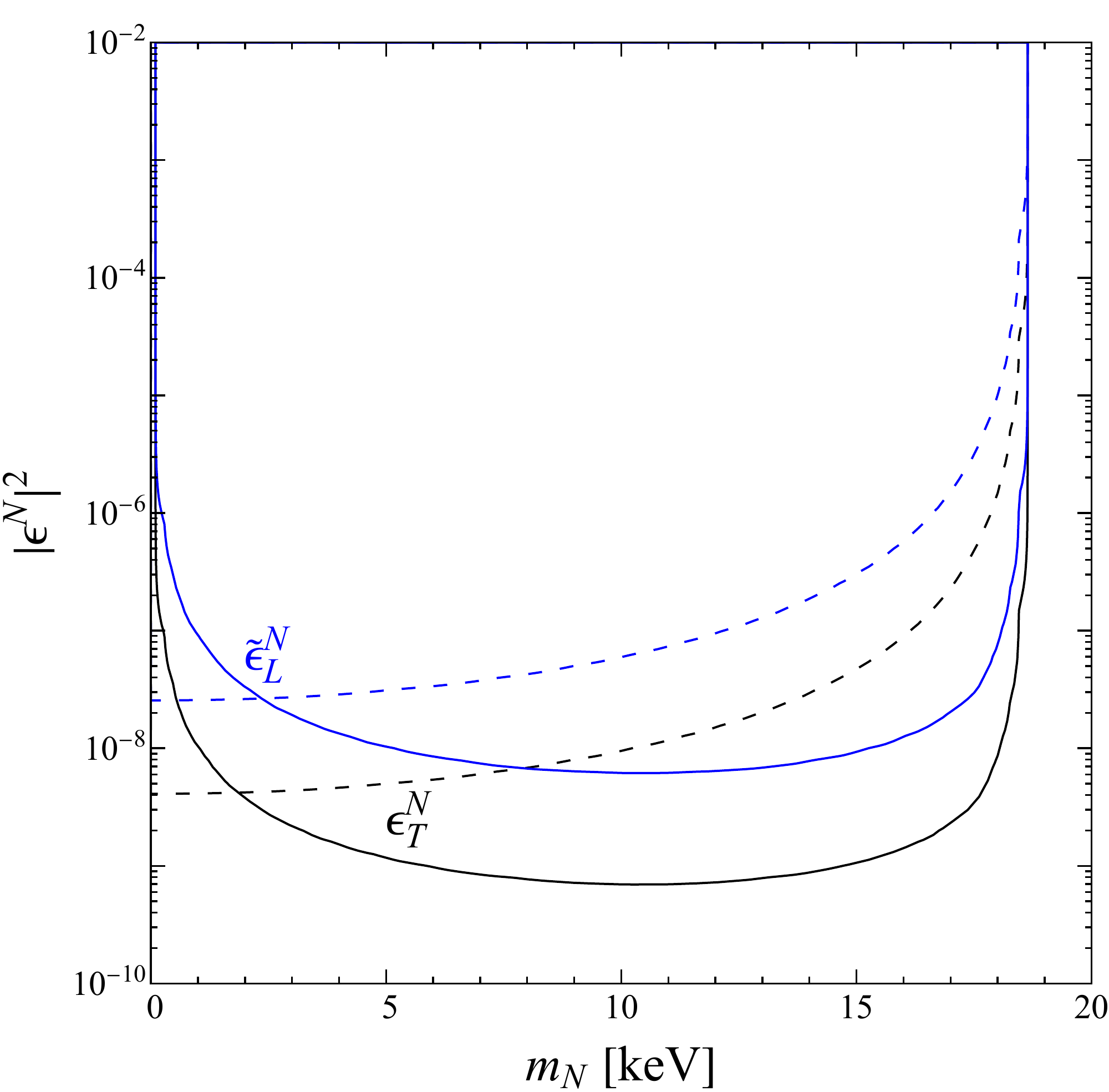}
	\caption{Sensitivity at $95\%$ CL to tensor and right-handed leptonic exotic currents as a function of the sterile neutrino mass using the energy (solid) and angular (dashed) distributions for currents parametrized by $\epsilon_T^N$ (black) and $\widetilde{\epsilon}_L^N$ (blue). The active-sterile mixing is set to zero, $\abs{V_{eN}}^2 = 0$.}
    \label{fig:SteCur}
\end{figure}
One can also look at the sensitivity to exotic currents of sterile neutrino beyond $V-A$. The sensitivity on these currents is dependent upon the mass of the sterile neutrino as can be seen in Fig.~\ref{fig:SteCur}. For the same reason as above, the energy spectrum is less sensitive for low and high masses. This does not apply however to the angular spectrum for low masses. Unlike with the active-sterile mixing, there is no multiplier of the active spectrum suppressing it as the sterile spectrum increases. This means that the additional currents contribute more significantly to changes in the rate. The normalisation uncertainty can hide this effect but not simultaneously to the angular change which varies differently and thus the angular spectrum does not lose sensitivity from this uncertainty. However, for heavier masses the reduced sensitivity instead comes from the sterile neutrino becoming kinematically impossible to produce and thus the impact of its presence is lost. 

\begin{table}[t!]
	\centering
	\begin{tabular}{c|cc} 
		\hline
		Parameter              & Energy           & Angular  \\
		\hline
		$\epsilon^N_S$         & $2\times10^{-4}$ & $2\times10^{-3}$ \\
		$\tilde{\epsilon}_S^N$ & $2\times10^{-4}$ & $2\times10^{-3}$\\
		$\epsilon_P^N$         & $5\times10^{-2}$ & $0.4$ \\
		$\tilde{\epsilon}_P^N$ & $5\times10^{-2}$ & $0.4$ \\
		$\epsilon_T^N$         & $3\times10^{-5}$ &$1\times10^{-4}$\\
		$\tilde{\epsilon}_T^N$ & $3\times10^{-5}$ & $1\times10^{-4}$ \\
		$\epsilon_L^N$         & $8\times10^{-5}$ & $2\times10^{-3}$ \\
		$\tilde{\epsilon}_L^N$ & $8\times10^{-5}$ & $2\times10^{-4}$\\
		$\epsilon_R^N$         & $8\times10^{-5}$ & $3\times10^{-4}$\\
		$\tilde{\epsilon}_R^N$ & $8\times10^{-5}$ & $5\times10^{-4}$\\
		\hline
	\end{tabular}
	\caption{Projected sensitivity at 95\%~CL on the coupling constants of exotic currents of a sterile neutrino with mass $m_N = 10$~keV. The sensitivity assumes  $10^{18}$ events in either the energy or angular distribution.}
	\label{table:Sterile}
\end{table}
Similarly, we consider the sensitivity to different currents with sterile neutrinos when considered individually with the sterile mass fixed to $10$~keV, i.e., the mass where the strongest sensitivity is approximately reached, see Fig.~\ref{fig:SteCur}. Unlike for the active case, there is no linear term in the coefficients as the sterile neutrino cannot interfere with the SM term meaning that turning on each current individually only leaves their quadratic terms. Furthermore, many of these terms actually have exactly or approximately identical quadratic terms. As in the active-sterile mixing case but unlike the active neutrino exotic currents, for a $10$~keV neutrino the energy distribution is more sensitive than the angular distribution. This is because the energy distribution contains the distinctive kink signature coming from the kinematic exclusion of heavy neutrinos above their threshold which is absent from the angular distribution by integrating over the entire spectrum. Table~\ref{table:Sterile} gives the $95\%$~CL upper bounds on the relevant coupling constants, for a 10~keV sterile neutrino. In general, the right-handed leptonic currents are more sensitive than their active counterparts whilst the left-handed currents are less sensitive. This is because the lack of a linear term for the left-handed currents makes the dependence quadratic and thus the current sensitivity is reduced; for the right-handed currents the active counterparts lacked a linear term anyway so instead the distinct spectrum from the sterile neutrino gives a greater sensitivity than the nearly identical extra terms in the active case. Again these parameters are taken to be real and positive, if seen as complex then the limits are equivalently placed upon $|\epsilon^N_X|$. 

\begin{figure}
	\centering
	\includegraphics[width=0.7\textwidth]{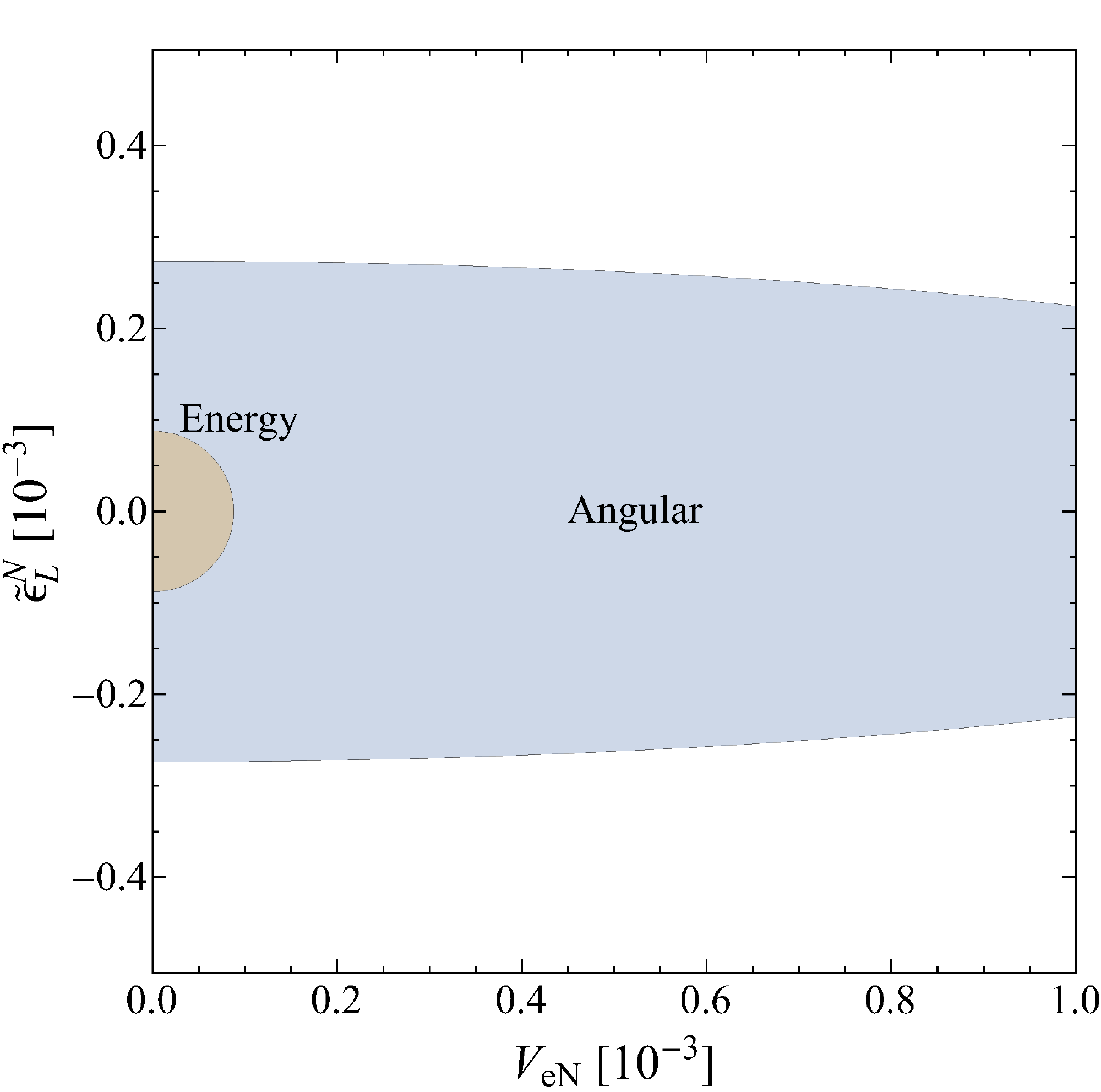}
	\caption{Projected sensitivity at $95\%$~CL on the active-sterile mixing $V_{eN}$ and the right-handed current of sterile neutrinos, $\tilde{\epsilon}^N_L$ for $m_N = 10$~keV and $10^{18}$ events. The region derived from the energy (angular) distribution is given in orange (blue).}
	\label{fig:ExoMix}
\end{figure}
Finally, we consider the presence of both an exotic current and the active-sterile mixing simultaneously, assuming a 10~keV sterile neutrino. The projected sensitivity at $95\%$ CL on $|V_{eN}|^2$ and the right-handed coupling $\tilde{\epsilon}^N_L$ can be seen in Fig.~\ref{fig:ExoMix}. Using the energy distribution, the sensitivity to either is practically identical, leading to a circular contour, because the primary sensitivity is to the kink which affects both equally. However, using the angular distribution, the sensitivity to the right-handed current is much higher because the corresponding contribution has an overall factor of $2g_A^2 + 2g_Ag_V$, unlike the active neutrino case (which the active-sterile mixing will enhance), which is proportional to $2g_A^2 - 2g_Ag_V$. Overall the angular sensitivity is weaker for aforementioned reasons.

\section{Conclusions}
\label{sec:conclusions}
Tritium $\beta$-decay is the method of choice for measuring the masses of active neutrinos model-independently. Its sensitivity is not affected by whether the neutrinos are of Dirac or Majorana nature, and as a laboratory measurement it is independent of astrophysical and cosmological considerations. Achieving sensitivities of order $\mathcal{O}(10~\text{meV})$, as would be required in the most pessimistic scenario of normally ordered neutrino states with a massless lightest neutrino, is nevertheless extremely challenging. The next generation of experiments, such as Project 8 and CRESDA, promise high resolutions near the endpoint with the potential for measurements across the entire $\beta$-spectrum. Their attempts to measure the neutrino masses also opens up the possibility to search for New Physics such as sterile neutrinos and exotic currents, complementing dedicated probes of the overall tritium spectrum (e.g., the proposed TRISTAN extension of the KATRIN experiment) and in other $\beta$-decaying or electron-capture isotopes, such as the proposed HUNTER experiment.

Current $\beta$-decay limits on the active-sterile mixing are of the order $|V_{eN}|^2 \sim 10^{-2}-10^{-3}$ coming from experiments with tritium or nickel. In this work we have analysed the prospect of future tritium experiments to probe the existence of sterile keV-scale neutrinos as well as exotic currents beyond $V-A$. The signature of emission of sterile neutrinos through active-sterile mixing is a kink in the spectrum with a reduced overall rate compared to the SM prediction. Assuming a statistics with $N_\text{tot} = 10^{18}$ events, corresponding to a sensitivity to the inversely ordered active neutrino scenario in the final eV below the endpoint, future experiments could reach sensitivities of $|V_{eN}|^2 \sim10^{-7}-10^{-8}$ within the mass range $1~\text{keV} \lesssim m_N \lesssim 18~\text{keV}$. 

Exotic charged currents beyond the Fermi interaction and its $V-A$ structure can also be probed in tritium decay, leaving an imprint on both the energy and angular distributions. The latter requires polarisation of the source and tritium is generally not ideal due to the low $Q$ value and correspondingly small kinetic energy of the electron. We nevertheless analyse it to assess the potential, e.g. in CRES type experiments. We have calculated the double differential decay rate in terms of both the electron energy and the electron momentum direction with respect to the nuclear spin of the tritium nucleus. The potential sensitivity to exotic currents varies significantly depending on the functional form of their contribution to the spectrum, with a suitable future tritium decay experiment expected to improve on some of the current constraints. For sterile neutrinos, which may also participate in exotic currents (maybe more likely so), the sensitivity depends on their mass. The strongest sensitivity is typically reached for $m_N \approx 10~\text{keV}$. Our analysis shows that future tritium $\beta$-decay experiments have the potential for many uses beyond measuring the active neutrino masses. Designing an experiment to observe the endpoint and entire spectrum with sufficient resolution as well as polarising the tritium and measuring the electron direction may be challenging but our work is intended as motivation to explore the prospects further.

\acknowledgments{The authors acknowledge support from the Science and Technology Facilities Council, part of U.K. Research and Innovation, Grant No. ST/T000880/1. The authors would like to thank the members of the QTNM collaboration for useful comments. The work of F.F.D. was performed in part at the Aspen Center for Physics, which is supported by National Science Foundation grant PHY-1607611. This work was partially supported by a grant from the Simons Foundation.}

\appendix

\section{Correction Factors to the Beta-Decay Spectrum}
\label{App:Cor}
In deriving the $\beta$-decay spectrum in Eq.~\eqref{eq:FulBet}, the factor $C(E_e)$ was added to account for additional theoretical corrections to the distribution of the electron kinetic energy. These corrections come from many different effects and are reproduced here from \cite{Kleesiek2019} and \cite{Mertens:2014nha}. We give a brief description of their origin along with their functional form for the sake of convenience. The impact of each correction on the bulk region of the distribution is discussed. They are approximately ordered by the magnitude of their effect upon the spectrum. The corrections are presented using the following conventions. Natural units ($\hbar = c = 1$) are used throughout unless otherwise stated. Energies and momenta are often given in units of the electron mass $m_e$, namely, $W = E_e/m_e$ is the electron energy in units of $m_e$, $W_0 = (E_e^\text{max} - V_f)/m_e$ is the tritium endpoint energy in units of $m_e$, with $V_f$ describing a change in the endpoint for excited states, and $k = \sqrt{W^2-1}$ is the magnitude of the electron spatial momentum in units of $m_e$. The electron speed in units of $c$ is $\beta = |\boldsymbol{p}_e| / E_e$. The atomic charge of the daughter nucleus, i.e., helium in our case, is denoted as $Z = 2$. The fine structure constant is $\alpha = 1/137.036$ and $\lambda_t = |g_A/g_V| = 1.247$ is the ratio of nuclear axial vector and vector coupling constants. The nuclear radius is denoted as $R_n \approx 2.88\times 10^{-3}/m_e$ \cite{Kleesiek2019}.

\begin{figure}[t!]
	\centering
	\includegraphics[width=\textwidth]{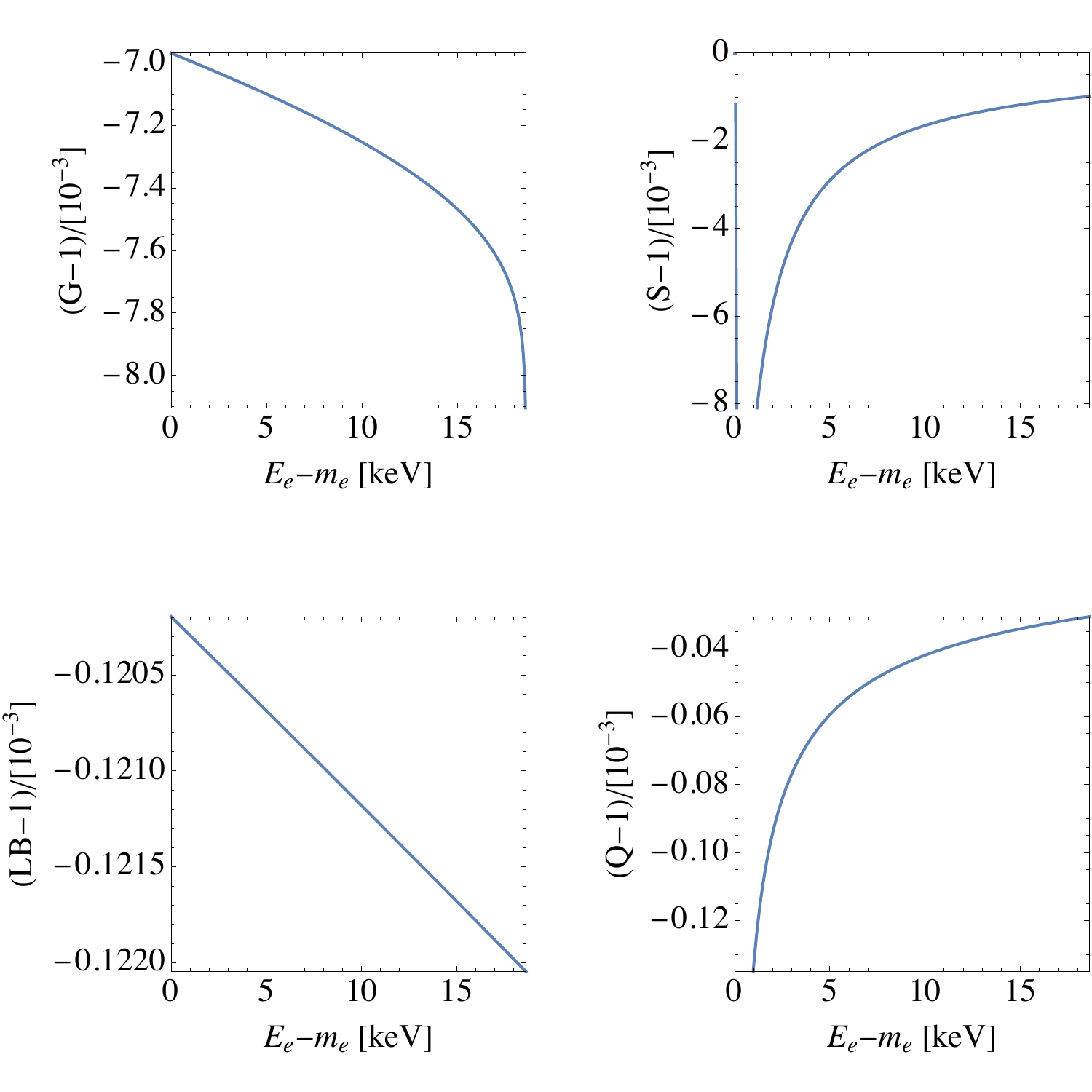}
	\caption{Plots showing the effect of theoretical corrections on the differential spectrum. Upper left: radiative corrections. Upper right: orbital electron shielding. Lower left: scaling of the electric field and convolution of electron and neutrino wave functions, both within the nucleus. Lower right: recoiling nuclear charge.}
	\label{fig:Corrections}
\end{figure}

\subsection{Fermi function}
The Fermi function corrects for the interaction between the charges of the emitted electron and the daughter nucleus. The fully relativistic expression of the Fermi function, $F(Z,E_e)$, can be derived from the relativistic eigenfunctions for hydrogen-like nuclei under the assumption that the electric field within the nucleus is very similar to that at its surface \cite{Fermi1934}. The functional form for $F(Z,E_e)$ is given by \cite{Kleesiek2019}
\begin{align}
	F(Z,E_e) = 
	4(2|\boldsymbol{p}_e|R_n)^{-2(1-\gamma)}
	e^{\pi\eta}
	\frac{|\Gamma(\gamma+i\eta)|^2}{\Gamma(2\gamma+1)^2}
	\approx
	\frac{2\pi\eta(1.002037-0.001427|\boldsymbol{p}_e|/E_e)}{1-e^{-2\pi\eta}},
\end{align}
where $\eta = \alpha Z E_e/|\boldsymbol{p}_e|$ is known as the Sommerfeld parameter and $\gamma = (1-(\alpha Z)^2)^{\frac{1}{2}}$. The approximate form (known as the Simpson approximation \cite{PhysRevD.23.649}) applies in the limit $\gamma\approx 1$, i.e., small $Z$ as is the case for tritium.

\subsection{Radiative Corrections}
Radiative corrections come from loop photon diagrams with interactions between the electron and either the initial neutron, final proton or intermediary $W$ boson. First order calculations lead to an infrared divergence near the endpoint due to the emission of soft (i.e., low energy), real photons. These can be corrected by considering soft photon emission to all orders. Taking this into consideration gives
\begin{align}
	G(E_e) &= 
	\left(\frac{E_e^\text{max}-E_e}{m_e}\right)^{(2\alpha/\pi) t(\beta)}
	\left\{1+\frac{2\alpha}{\pi}\left[t(\beta)
	\left(\ln 2-\frac{3}{2}+\frac{E_e^\text{max}-E_e}{E_e}\right) \right. \right. \nonumber\\
	&+ \frac{1}{4}(t(\beta)+1)\left(2(1+\beta^2) - 2\ln\left(\frac{2}{1-\beta}\right) + \frac{(E_e^\text{max}-E_e)^2}{6E_e^2}\right) \nonumber\\
	&\left. \left. + \frac{1}{2\beta}
	\left(L(\beta) - L(-\beta) + L\left(\frac{2\beta}{1+\beta}\right) 
	+ \frac{1}{2}L\left(\frac{1-\beta}{2}\right) 
	- \frac{1}{2}L\left(\frac{1+\beta}{2}\right)\right)\right]\right\},
\end{align}
where $t(\beta) = \text{arctanh}(\beta)/\beta-1$ and $L(x)$ is Spence's function, $L(x) = \int_0^x \frac{\ln(1-t)}{t}dt$ \cite{Repko:1984cs}.

\subsection{Screening}
The Coulomb interaction between the emitted electron and the daughter nucleus assumes that the nuclear charge is bare. Therefore, an extra correction factor $S(Z, E_e)$ needs to be considered in order to take into account the screening effect of the bound atomic electron on the Coulomb field. This correction is of the form \cite{Behrens1982-zb}
\begin{align}
	S(Z,W) = 
	\frac{\overline{W}}{W}\left(\frac{\overline{k}}{k}\right)^{-1+2\gamma}
	\frac{\abs{\Gamma(\gamma+i\overline{\eta})^2}}{\abs{\Gamma(\gamma+i\eta)^2}}
	e^{\pi(\overline{\eta}-\eta)},
\end{align}
where $\overline{W} = W - V_0/m_e$, $\overline{k} = \sqrt{\overline{W}^2-1}$ and $\overline{\eta} = \alpha Z \overline{W}/\overline{k}$. Here, $V_0 = (76\pm 10)$~eV is the nuclear screening potential of a $1s$ helium orbital electron \cite{PhysRevC.60.034608}.

\subsection{Recoil Effects}
Recoil effects arise from the weak magnetism and $V-A$ terms, and the three-body phase space. These terms are all dependent upon the $Q$ value and thus vanish in the zero recoil case. To take them into account, usually a correction term $R(W,W_0)$ is applied to the energy distribution,
\begin{align}
	R(W, W_0) = 1 + \frac{1}{C}\left(AW - \frac{B}{W}\right),
\end{align}
where
\begin{align}
	A &= 2(5\lambda_t^2+\lambda_t\mu+1)/m_\text{He}, \nonumber\\
	B &= 2\lambda_t(\mu+\lambda_t)/m_\text{He}, \\
	C &= 1 + 3\lambda_t^2 - bW_0, \nonumber
\end{align}
and $\mu = 5.107$ is the difference in bare triton and helion magnetic moments. In this work the differential decay rates are calculated from the full matrix element, and hence include all significant $V-A$ terms and also take into account 3-body effects. Therefore, this correction term is not part of the overall correction factor, but is nevertheless included here for completeness. 

\subsection{Finite Nuclear Size}
In many cases it is sufficient to consider the nucleus as a point-like structure. However, the finite structure of the daughter nucleus can be included by using the corrections $L(Z,W)$ and $B(Z,W)$ \cite{wilkinson1990evaluation}. The function $L(Z,W)$ accounts for the scaling of the Coulomb field and is given by
\begin{align}
	L(Z,W) = 
	1 + \frac{13}{60}(\alpha Z)^2 
	- \frac{W R_n\alpha Z}{15}\frac{41-26\gamma}{2\gamma -1}
	- \frac{\alpha Z R_n\gamma}{30W}\frac{17-2\gamma}{2\gamma -1}.
\end{align}
The function $B(Z,W)$, on the other hand, takes into account the convolution of the electron and neutrino wave functions with that of the nucleus for the whole nuclear volume,
\begin{align}
	B(Z,W) = 1 + B_0 + B_1 W + B_2 W^2,
\end{align}
with
\begin{align}
	B_0 &= -\frac{233}{630}(\alpha Z)^2-\frac{1}{5}(W_0 R_n)^2 + \frac{2}{35}(W_0 R_n\alpha Z), \nonumber\\
	B_1 &= -\frac{21}{35}R_n\alpha Z+\frac{4}{9} W_0 R_n^2, \\
	B_2 &= -\frac{4}{9}R_n^2. \nonumber
\end{align}

\subsection{Recoiling Coulomb Field}
The Fermi function treats the nucleus as infinitely heavy. In reality, it is not and as a result the nucleus recoils upon the emission of the electron and neutrino. This in turn means the origin of the Coulomb field is in fact a moving source - an effect captured by the function~\cite{wilkinson1982analysis}
\begin{align}
	Q(Z, W, W_0) = 
	1 - \frac{\pi\alpha Z}{m_\text{He} k}
	\left(1 + \frac{1 - \lambda_t^2}{1 + 3\lambda_t^2}\frac{W_0-W}{3W}\right).
\end{align}

\subsection{Orbital Electron Interactions}
The bound orbital electron will undergo a sudden change of state from a hydrogen energy level to a helium energy level. It may also interact with the emitted $\beta$-electron, with the possibility that the emitted electron may knock the atomic electron out of the atom, and become bound, while the original orbital-electron is ejected and is observed as the $\beta$-particle. Considering both of these scenarios for decay to the $1s$ helium state induces the correction~\cite{Haxton:1985cz}
\begin{equation}
	I(Z,W) = 1 + \frac{729}{256}a^2(\tau) + \frac{27}{16}a(\tau),
\end{equation}
with
\begin{align}
\label{eq:atau}
	a(\tau) = 
	\exp\left[2\tau\arctan\left(-\frac{2}{\tau}\right)\right]
	\left(\frac{\tau^2}{1+\frac{1}{4}\tau^2}\right)^2,
\end{align}
and $\tau = -2\alpha / k$. This case only accounts for exchange with the $1s$ electron and thus $I(Z,W)$ is not included in our calculation, as the more exact continuous emission correction described in the next section is applied instead.

\subsection{Excited States and Continuous Emission of The Atomic Electron}
\label{subsec:cont}
The tritium orbital electron is in principle indistinguishable from the emitted electron after $\beta$-decay. Thus it is important to consider what happens to this electron which may either end up in a bound helium state or in the continuum. Included in this must also be the possibility of quantum mechanical exchange with the emitted electron becoming bound and vice versa. The energy of the orbital electron determines the energy available to the $\beta$-electron as it may give off or take energy from the decay process. Table~\ref{table:emission}, based on \cite{PhysRevC.27.1815}, summarises the energy gap between the possible excitation states of the atomic electron left behind, and also states the transition probabilities, i.e., branching ratio, for each.

The probability of population of each excited final state is shown in the zeroth-order sudden approximation (third column) and includes the interaction between the $\beta$ and atomic electron. The sudden approximation, which is well supported in this 4-point interaction approximation, assumes that a rapid change in the Hamiltonian allows a transition amplitude of the form \cite{PhysRevC.27.1815} $\bra{b}U(t_2, t_1)\ket{a}$, to be simply approximated by the overlap $\bra{b}\ket{a}$, as $t_2 - t_1 \sim 0$, where $U(t_2,t_1)$ is some transition function depending on time $t_1$ and a later time $t_2$. Therefore, the transition probability is $|\bra{b}U(t_2,t_1)\ket{a}|^2\sim |\bra{b}\ket{a}|^2$. 

The total decay rate is thus expressed as a sum over decays to different energy levels. The decay rate to an individual energy level is identical as described above, however modified by a shift in the endpoint. This is done by shifting the effective masses for the tritium and helium by including the electron and its binding or kinetic energy,
\begin{align}
	E_e^\text{max}\to \begin{cases} 
		E_e^\text{max}(n), \; m_\text{H}\to m_\text{H}-R_h, \; 
		m_\text{He}\to m_\text{He} + m_e - 4R_h/n^2, & n = 1,2,3,\dots, \\
		E_e^\text{max}(\tau), \; m_\text{H} \to m_\text{H} - R_h, \; 
		m_\text{He} \to m_\text{He} + m_e + 4R_h/\tau^2, & \tau = -2\alpha/k,
	\end{cases}
\end{align}
where $R_h=13.61$~eV is the Rydberg energy \cite{PhysRevD.98.030001} and with the top case being to a helium bound state and the bottom being to the continuum. Combining these terms with the appropriate weightings gives the full decay rate over all possible bound and continuum helium final states \cite{Haxton:1985cz},
\begin{align}
\label{eq:OrbitSum}
	\frac{d\Gamma_\text{full}}{dE_e} &= 
	\sum_{n=1}^{\infty}2\frac{d\Gamma}{dE_e}(E_e^\text{max}(n))
	\left[256n^5\frac{(n-2)^{2n-4}}{(n+2)^{2n+4}} + \frac{\alpha^2(\tau)}{n^3} 
	- 16n\frac{(n-2)^{n-2}}{(n+2)^{n+2}}\alpha(\tau)\right] \nonumber\\
	&+ \frac{1}{\pi}\int_{-\infty}^{\tau}\frac{d\tau'}{{\tau'}^4}
	\frac{d\Gamma}{dE_e}(E_e^\text{max}(\tau'))
	\frac{2\pi\tau'}{e^{2\pi\tau'}-1}
	\left[\alpha^2(\tau) - \alpha(\tau)\alpha(\tau') + \alpha^2(\tau')\right],
\end{align}
Here, $a(\tau)$ is given in Eq.~\eqref{eq:atau} and $(n-2)^{n-2}\to 1$ as $n\to 2$ for the sake of evaluation. The first line involves summing over decays to discrete bound energy levels while the second is an integral over all possible continuum energies for the non-measured electron. This leads to an energy dependence for the probability of decay to a particular energy level as can be seen in Fig.~\ref{fig:EnLev}.

\section{Full Expressions for Tritium Differential Decay Rate} 
\label{App:full-expressions}

The $\beta$-decay rate can be generally written in the form
\begin{align}
\label{eq:AppDiffRate}
	\frac{d\Gamma}{dE_e d\Omega_e} &= a(E_e) + b(E_e)\cos\theta_e
\end{align}
If two effective operators $\mathcal{O}_X$, $\mathcal{O}_Y$ in Eq.~\eqref{eq:FullLag} contribute to the decay for a given neutrino mass eigenstate, the transition matrix element may be composed as
\begin{align}
	T = \epsilon_X T_X + \epsilon_{Y} T_Y,
\end{align}
where $T_X$ and $T_Y$ are the individual matrix elements and $\epsilon_X$, $\epsilon_Y$ are the associated effective coupling constants. The terms in the differential decay rate Eq.~\eqref{eq:AppDiffRate} can then be written as
\begin{align}
	a(E_e) &= |\epsilon_X|^2 a_\text{X}(E_e) + \text{Re}(\epsilon_X\epsilon_Y^*) a_{X,Y}(E_e) 
	+ |\epsilon_Y|^2 a_Y(E_e), \\
	b(E_e) &= |\epsilon_X|^2 b_\text{X}(E_e) + \text{Re}(\epsilon_X\epsilon_Y^*) b_{X,Y}(E_e) 
	+ |\epsilon_Y|^2 b_Y(E_e).
\end{align}
Most importantly, as the SM contribution with both left-handed lepton and quark vector currents is always present, we consider its interference with an exotic operator. Thus, $X = LL$, $\epsilon_X = 1$ (for a single active neutrino) and $Y$ is one of the exotic operators in Eq.~\eqref{eq:ExL} or Eq.~\eqref{eq:ExLN}.

Below are presented the exact expressions for some of these terms. We concentrate on the individual contribution of each operator and the interference of exotic operators with that in the SM. The expressions are written in terms of a single generic neutrino mass eigenstate $\nu$ with mass $m_\nu$, which can be either one of the active neutrinos, $\nu_{1,2,3}$, or a sterile neutrino $\nu_N$. In the latter case, there is no interference with any contributions from active neutrinos. For simplification, we define
\begin{align}
	\widetilde{C}(E_e) &= \frac{1}{4\pi}\frac{G_F^2|V_{ud}|^2}{2\pi^3} C(E_e) 
	                       \Theta(E_e^\text{max} - E_e - m_\nu), \\
	m_{12}^2           &= (p_\text{H} - p_e)^2 = m_\text{H}^2 - 2m_\text{H}E_e + m_e^2, \\
	|\boldsymbol{p}_e| &= \sqrt{E_e^2 - m_e^2}, \\
	\mu                &= (m_\nu + m_\text{He})/m_\text{H}, \\
	y                  &= E_e^\text{max} - E_e, \\
	\widetilde{y}      &= \sqrt{y\left(y + m_\nu \frac{2m_\text{He}}{m_\text{H}}\right)} \\
	\alpha&=m_{\text{H}}E_e^{\rm{max}}+m_{\nu}^2+m_{\text{He}}m_{\nu}.
\end{align}

\subsection{Individual SM and Exotic Contributions}
\textbf{Vector Currents} $LL$, $RR$, $RL$, $LR$

The SM contribution $(LL)$ is related to that of $\tilde\epsilon_R$ $(RR)$ by using $\gamma_5 \to -\gamma_5$ (or equivalently $S \to -S$). This results in the same energy distributions and reverse angular correlations,
\begin{align}
	a_{LL}(E_e) &= a_{RR}(E_e) \nonumber\\
	&= \widetilde {C}(E_e) \frac{m_\text{H}^2|\boldsymbol{p}_e|}{m_{12}^2} \widetilde{y}
	\nonumber\\ 
	&\times\left\{(g_V\!+\!g_A)^2 \left[\frac{m_\text{H}(m_\text{H}-E_e)}{m_{12}^2}\frac{m_\text{H}E_e-m_e^2}{m_{12}^2}
	(y\!+\!\mu m_\nu)
	(y\!+\!\mu m_\text{He})\right. 
	-\frac{m^2_\text{H}|\boldsymbol{p}_e|^2}{3m_{12}^4} \widetilde{y}^2\right] \nonumber\\
	&\quad+(g_V\!-\!g_A)^2E_e\left(y+m_\nu\frac{m_\text{He}}{m_\text{H}}\right)
	+(g_A^2\!-\!g_V^2)m_\text{He}\frac{m_\text{H}E_e-m_e^2}{m_{12}^2}(y+\mu m_\nu)  \Biggr\}, \\
	b_{LL}(E_e) &= - b_{RR}(E_e) \nonumber\\
	&= -\widetilde{C}(E_e)\frac{m_\text{H}}{m_{12}^2}|\boldsymbol{p}_e|^2
	\widetilde{y} \nonumber\\
	&\times\left\{\left[(g_A-g_V)^2 m_\text{H} 
	+ (g_A^2-g_V^2)m_\text{He}\frac{m_\text{H}(m_\text{H}-E_e)}{m_{12}^2}\right.\right. \nonumber\\
	&\quad\,\,\,+(g_A^2-g_V^2)\frac{m_\text{H}m_\text{He}}{m_{12}^2}E_e 
	+(g_A+g_V)^2\frac{m_\text{H}}{m_{12}^2}(\alpha-m_e^2)\nonumber\\ 
	&\quad\,\,\,\left.-(g_A+g_V)^2\frac{m_\text{H}^2}{m_{12}^2}(y+\mu m_\nu)
	\frac{m_\text{H}(m_\text{H}-E_e)}{m_{12}^2}\right]
	(y+\mu m_\nu)\nonumber\\
	&\quad\,\,\,\left.-(g_A-g_V)^2m_\nu^2 
	-\frac{1}{3}(g_A+g_V)^2\frac{m_\text{H}^3(m_\text{H}-E_e)}{m_{12}^4}\widetilde{y}^2\right\}. 
\end{align}

The contributions related to $\tilde\epsilon_L$ $(LR)$ and $\epsilon_R$ $(RL)$ can be obtained from the SM $(LL)$ terms by applying $g_A \to -g_A$ with an additional $S \to -S$ for the right-handed lepton term,
\begin{align}
	a_{LR}(E_e) &= a_{RL}(E_e) = a_{LL}(E_e)(g_A \to -g_A), \\
	b_{LR}(E_e) &= -b_{RL}(E_e) = -b_{LL}(E_e)(g_A \to -g_A).
\end{align}
This makes a negligible difference in form.

\paragraph{Scalar Currents}
\begin{align}
	a_S(E_e) &= \frac{1}{2}g_S^2\widetilde {C}(E_e) \frac{m_\text{H}^2|\boldsymbol{p}_e|}{m_{12}^2} \widetilde{y}
	\nonumber\\ 
	&\times\Biggr\{(m_{\text{He}}+m_{\text{H}}-E_e)\frac{m_{\text{H}}E_e-m_e^2}{m_{12}^2}(y\!+\!\mu m_{\nu})-\frac{1}{3}\frac{m_{\text{H}}^2|\boldsymbol{p}_e|^2}{m_{12}^4}\widetilde{y}^2 \nonumber\\
	&-\frac{m_{\text{H}}(m_{\text{H}}-E_e)}{m_{12}^2}\frac{m_{\text{H}}E_e-m_e^2}{m_{12}^2}(y\!+\!\mu m_{\nu})^2\Biggr\},  \\
	b_S(E_e) &= 0.
\end{align}
\begin{align}
	a_{\tilde S}(E_e) &= a_S(E_e), \\
	b_{\tilde S}(E_e) &= 0.
\end{align}

\paragraph{Pseudoscalar Currents}
\begin{align}
	a_P(E_e) &= a_S(E_e)(g_S \to g_P), \\
	b_P(E_e) &= 0.
\end{align}
\begin{align}
	a_{\tilde P}(E_e) &= a_S(E_e)(g_S \to g_P), \\
	b_{\tilde P}(E_e) &= 0.
\end{align}

\paragraph{Tensor Currents}
\begin{align}
	a_T(E_e) &= a_{\tilde T}(E_e) = \nonumber\\
	&=16 \widetilde{C}(E_e) g_T^2 \frac{m_\text{H}^2|\boldsymbol{p}_e|}{m_{12}^2} \widetilde{y}
	\Biggr[
    	\frac{m_{\text{H}}(m_{\text{H}}-E_e)}{m_{12}^2}\frac{m_{\text{H}}E_e-m_e^2}{m_{12}^2}(y\!+\!\mu m_{\nu})(3y+2\mu m_{\text{He}}+\mu m_{\nu}) \nonumber \\
    	&-\frac{1}{3}\frac{m_{\text{H}}^2|\boldsymbol{p}_e|^2}{m_{12}^4}\widetilde{y}^2+2E_e\left(y+m_{\nu} \frac{m_{\text{He}}}{m_{\text{H}}}\right)-2(m_{\text{H}}-E_e)(y\!+\!\mu m_{\nu})\frac{m_{\text{H}}E_e-m_e^2}{m_{12}^2}
	\Biggr], \\
	b_T(E_e) &= - b_{\tilde T}(E_e) \nonumber\\
	&= 16 \widetilde{C}(E_e)
	g_T^2\frac{m_\text{H}^2}{m_{12}^2}|\boldsymbol{p}_e|^2 \widetilde{y}
	\left[
		 \frac{1}{3}\frac{m_\text{H}(m_\text{H}-E_e)}{m_{12}^2}
		 \frac{m_\text{H}|\boldsymbol{p}_e|}{m_{12}^2} \widetilde{y} \right. \nonumber\\
		&\qquad\left.+\left(2+\frac{m_\text{H}E_e-m_e^2}{m_{12}^4}m_\text{H}(y+\mu m_\nu)
		- \frac{m_\text{H}E_e}{m_{12}^2}-\frac{2m_\nu^2}{m_\text{H}|\boldsymbol{p}_e|}\right)(y+\mu m_\nu)
	\right].
\end{align}
\hspace{10cm}

\subsection{Interference Terms}

\paragraph{Vector Currents: $(LL)$ with $(LR)$}
\begin{align}
	a_{LL,LR}(E_e) &= \widetilde{C}(E_e)m_e m_\nu\frac{m_\text{H}^2|\boldsymbol{p}_e|}{m_{12}^2} 
	\widetilde{y}
	\Bigg[2(g_A^2 - g_V^2)m_\text{He} + (g_A^2\!+\!g_V^2)(m_\text{H}-E_e) \nonumber\\
	&\qquad\qquad\qquad\left.- (g_A^2 + g_V^2)\frac{m_\text{H}(m_\text{H} - E_e)}{m_{12}^2}(y + \mu m_\nu)\right], \\
	b_{LL,LR}(E_e) &= 
	-\widetilde{C}(E_e) g_A g_V m_e m_\nu |\boldsymbol{p}_e|^2 \frac{m_\text{H}}{m_{12}^2} 
	\widetilde{y} 
	\left[1-\frac{m_\text{H}}{m_{12}^2}(y \!+\! \mu m_\nu)\right].
\end{align}
All other interference terms with differing leptonic handedness can be related by $\gamma^5 \to -\gamma^5$ (with $S \to -S$) or $g_A \to -g_A$  ,
\begin{align}
    a_{LR,LL}(E_e)&=a_{RR,RL}(E_e)=a_{RL,RR}(E_e)=a_{LL,LR}(E_e), \\
    b_{LR,LL}(E_e)&=b_{RR,RL}(E_e)=-b_{RL,RR}(E_e)=-b_{LL,LR}(E_e).
\end{align}

\paragraph{Vector Currents: $(LL)$ with $(RL)$}
\begin{align}
	a_{LL,RL}(E_e) &= -\widetilde{C}(E_e) \frac{m_\text{H}^2|\boldsymbol{p}_e|}{m_{12}^2} 
	\widetilde{y} \nonumber\\ 
	&\times\left\{(g_A^2\!-\!g_V^2) \left[\frac{m_\text{H}(m_\text{H}\!-\!E_e)}{m_{12}^2}
	\frac{m_\text{H}E_e\!-\!m_e^2}{m_{12}^2}(y\!+\!\mu m_\nu)(y\!+\!\mu m_\text{He})  
	\!-\! \frac{m_\text{H}^2|\boldsymbol{p}_e|^2}{3m_{12}^4}\widetilde{y}^2\right]\right.
	\nonumber \\
	&\quad\, \left.+ (g_A^2\!-\!g_V^2)E_e\left(ym_\nu\frac{m_\text{He}}{m_\text{H}}\right) 
	\!+\!(g_A^2\!+\!g_V^2)m_\text{He}\left(\frac{m_\text{H}E_e\!-\!m_e^2}{m_{12}^2}(y\!+\!\mu m_\nu)\right)\right\}, \\
	b_{LL,RL}(E_e) &= - \widetilde{C}(E_e) \frac{m_\text{H}}{m_{12}^2}|\boldsymbol{p}_e|^2 
	\widetilde{y} \nonumber\\
	&\times\left\{\left[(g_A^2-g_V^2)m_\text{H}+(g_A^2+g_V^2)m_\text{He}
	\frac{m_\text{H}(m_\text{H}-E_e)}{m_{12}^2} 
	+(g_A^2+g_V^2)\frac{m_\text{H}m_\text{He}}{m_{12}^2}E_e \right.\right. \nonumber\\ 
	&\quad\,\,\,\left.+(g_A^2\!-\!g_V^2)\frac{m_\text{H}}{m_{12}^2}(\alpha\!-\!m_e^2) 
	\!-\!(g_A^2\!-\!g_V^2)\frac{m_\text{H}^2}{m_{12}^2}
	(y\!+\!\mu m_\nu)\frac{m_\text{H}(m_\text{H}\!-\!E_e)}{m_{12}^2}\right]
	(y\!+\!\mu m_\nu) \nonumber\\
	&\quad\,\,\,\left.- (g_A^2-g_V^2)m_\nu^2 
	 - \frac{1}{3}(g_A^2-g_V^2)\frac{m_\text{H}^3(m_\text{H}-E_e)}{m_{12}^4}\widetilde{y}^2\right\}.
\end{align}
All other interference terms with differing hadronic handedness can be related by $\gamma^5 \to -\gamma^5$ (with $S \to -S$) or $g_A \to -g_A$  ,
\begin{align}
    a_{LR,RR}(E_e)&=a_{RR,LR}(E_e)=a_{RL,LL}(E_e)=a_{LL,RL}(E_e), \\
    b_{LR,RR}(E_e)&=b_{RR,LR}(E_e)=-b_{RL,LL}(E_e)=-b_{LL,RL}(E_e).
\end{align}

\paragraph{Vector Currents: $(LL)$ with $(RR)$}
\begin{align}
	a_{LL,RR}(E_e) &= -\widetilde{C}(E_e) m_e m_\nu \frac{m_\text{H}^2|\boldsymbol{p}_e|}{m_{12}^2} \widetilde{y}
	\Bigg[(2(g_A^2+g_V^2)m_\text{He} + (g_A^2-g_V^2)(m_\text{H}-E_e) \nonumber\\
	&\qquad\qquad\qquad\qquad\qquad\quad\,\,\,\,
	\left.- (g_A^2-g_V^2)\frac{m_\text{H}(m_\text{H}-E_e)}{m_{12}^2}(y + \mu m_\nu)\right], \\
	b_{LL,RR}(E_e) &= 0.
\end{align}
All other interference terms differing in handedness for both can be related by $\gamma^5 \to -\gamma^5$ (with $S \to -S$) or $g_A \to -g_A$  ,
\begin{align}
    a_{LR,RL}(E_e)&=a_{LL,RR}(E_e)=a_{RL,LR}(E_e)=a_{RR,LL}(E_e), \\
    b_{LR,RL}(E_e)&=b_{LL,RR}(E_e)=b_{RL,LR}(E_e)=b_{RR,LL}(E_e)=0.
\end{align}

\paragraph{$LL$ with Left-handed Scalar Current}
\begin{align}
	a_{LL,S}(E_e) &= \widetilde{C}(E_e)g_S g_V m_e
	\frac{m_\text{H}|\boldsymbol{p}_e|}{m_{12}^2}\widetilde{y} \nonumber\\ 
	&\times \left[m_\text{H}
	 \left(1+\frac{m_\text{He}(m_\text{H}-E_e)}{m_{12}^2}\right)
	 \left(y + \mu m_\nu\right) - m_\nu^2\right], \\
	b_{LL,S}(E_e) &= - \widetilde{C}(E_e) g_S g_A m_e 
	m_\text{He}\frac{m_\text{H}^2}{m_{12}^2}\frac{|\boldsymbol{p}_e|^2}{m_{12}^2}
	\widetilde{y}(y + \mu m_\nu).
\end{align}

\paragraph{$LL$ with Right-handed Scalar Current}
\begin{align}
	a_{LL,\tilde S}(E_e) &= - \widetilde{C}(E_e) g_S g_V 
	m_\nu \frac{m_\text{H}|\boldsymbol{p}_e|}{m_{12}^2} \widetilde{y} \nonumber\\ 
	&\times \left[(m_\text{H} + m_\text{He}) E_e - m_e^2 - m_\text{H}\frac{m_\text{H}E_e - m_e^2}{m_{12}^2}
	\left(y + \mu m_\nu\right)\right], \\
	b_{LL,\tilde S}(E_e) &= -\widetilde{C}(E_e) g_S g_A m_\nu
	M \frac{|\boldsymbol{p}_e|^2}{m_{12}^2} \widetilde{y}  
	\left[\frac{m_\text{H}^2}{m_{12}^2}(y + \mu m_\nu) - (m_\text{H}+m_\text{He})\right].
\end{align}

\paragraph{$LL$ with Left-handed Pseudoscalar Current}
\begin{align}
	a_{LL,P}(E_e) &= \widetilde{C}(E_e) g_P g_A 
	m_e \frac{m_\text{H}|\boldsymbol{p}_e|}{m_{12}^2} \widetilde{y} \nonumber\\ 
	&\times \left[m_\text{H}\left(\frac{m_\text{He}(m_\text{H}-E_e)}{m_{12}^2} - 1\right)
	\left(y + \mu m_\nu\right) + m_\nu^2\right], \\
	b_{LL,P}(E_e) &= - \widetilde{C}(E_e) g_P g_V m_e 
	m_\text{He} \frac{m_\text{H}^2}{m_{12}^2}\frac{|\boldsymbol{p}_e|^2}{m_{12}^2}\widetilde{y}
	(y + \mu m_\nu).
\end{align}

\paragraph{$LL$ with Right-handed Pseudoscalar Current}
\begin{align}
	a_{LL,\tilde P}(E_e) &= \widetilde{C}(E_e) g_P g_A
	m_\nu \frac{m_\text{H}|\boldsymbol{p}_e|}{m_{12}^2} \widetilde{y} \nonumber\\ 
	&\times \left[(m_\text{H}-m_\text{He})E_e - m_e^2 - m_\text{H}\frac{m_\text{He}E_e - m_e^2}{m_{12}^2}
	(y+\mu m_\nu)\right], \\
	b_{LL,\tilde P}(E_e) &= \widetilde{C}(E_e) g_P g_V m_\nu 
	m_\text{H}\frac{|\boldsymbol{p}_e|^2}{m_{12}^2} \widetilde{y} 
	\left[\frac{m_\text{H}^2}{m_{12}^2}(y + \mu m_\nu) - (m_\text{H} - m_\text{He})\right].
\end{align}

\paragraph{$LL$ with Left-handed Tensor Current}
\begin{align}
	a_{LL,T}(E_e) &= -12 \widetilde{C}(E_e) g_T 
	m_e \frac{m_\text{H}|\boldsymbol{p}_e|}{m_{12}^2}\widetilde{y} \nonumber \\ 
	&\times\left[m_\text{H}\left((g_A+g_V)\frac{m_\text{He}(m_\text{H}-E_e)}{m_{12}^2}+(g_A-g_V)\right)
	 (y+\mu m_\nu) - (g_A-g_V)m_\nu^2\right], \\
	b_{LL,T}(E_e) &=  12\widetilde{C}(E_e) g_T (g_A+g_V) m_e 
	m_\text{He}\frac{m_\text{H}^2}{m_{12}^2}\frac{|\boldsymbol{p}_e|^2}{m_{12}^2}\widetilde{y}
	(y + \mu m_\nu).
\end{align}

\paragraph{$LL$ with Right-handed Tensor Current}
\begin{align}
	a_{LL,\tilde T}(E_e) &= 12\widetilde{C}(E_e) g_T
	m_\nu \frac{M|\vec{\boldsymbol{p}}_e|}{m_{12}^2}\widetilde{y} \nonumber \\ 
	&\times\left[(g_A\!+\!g_V)\left(m_\text{H}E_e - m_e^2 
	- m_\text{H}\frac{m_\text{H}E_e-m_e^2}{m_{12}^2}(y + \mu m_\nu)\right) 
	\!+\! (g_A\!-\!g_V)m_\text{He}E_e\right], \\
	b_{LL,\tilde T}(E_e) &= 4 \widetilde{C}(E_e) g_T m_\nu m_\text{H}\frac{|\boldsymbol{p}_e|^2}{m_{12}^2} \widetilde{y} \nonumber \\ 
	&\times\left[3(g_A-g_V)m_\text{He} + (g_A+g_V)m_\text{H} 
	- (g_A+g_V)\frac{m_\text{H}^2}{m_{12}^2}(y+\mu m_\nu)\right].
\end{align}

\bibliographystyle{JHEP}
\bibliography{bibliography.bib}

\end{document}